\documentclass{aiaa-tc}

\def\softd{{\leavevmode\setbox1=\hbox{d}%
          \hbox to 1.05\wd1{d\kern-0.4ex{\char039}\hss}}}

 \usepackage{wrapfig}
 \usepackage{threeparttable}
 \usepackage{dcolumn}
  \newcolumntype{d}{D{.}{.}{-1}}
 \usepackage{nomencl}
  \makeglossary
 \usepackage{subfigure}
 \usepackage{fancyvrb}
  \fvset{fontsize=\footnotesize,xleftmargin=2em}
 \usepackage{lettrine}
 \usepackage[dvips]{dropping}
 \usepackage{hyperref}
 \usepackage{amsmath} 
\usepackage{graphicx}
\graphicspath{ {figures/} }
\usepackage{float}
 \newcommand{\nin}{\noindent} 
\newcommand{\be} {\begin{equation}}
\newcommand{\ee} {\end{equation}}
\newcommand{\bea} {\begin{eqnarray}}
\newcommand{\eea} {\end{eqnarray}} 
\newcommand{\ba} {\begin{array}}
\newcommand{\ea} {\end{array}} 

\newcommand{\fr}{\displaystyle \frac} 

\newcommand{\bi}{\begin{itemize}} 
\newcommand{\ei}{\end{itemize}}

 \title{A Lattice Boltzmann Relaxation Scheme for Inviscid Compressible Flows}

  \author{
  S.V. Raghurama Rao\thanks{Associate Professor (Corresponding Author, email: raghu@aero.iisc.ernet.in)},
  Rohan Deshmukh\thanks{PhD student; Indian Institute of Science, Bangalore}
  \ and
  Sourabh Kotnala\thanks{MSc(Engg.) student; Indian Institute of Science, Bangalore}\\
  {\normalsize\itshape
   Department of Aerospace Engineering, 
   Indian Institute of Science, 
   Bangalore-560012, India
 }}

 \AIAApapernumber{200?-????}
 \AIAAconference{AIAA CFD Conference, 24-27 June, San Diego, California}
 \AIAAcopyright{\AIAAcopyrightD{200?}}

\begin{document}

\maketitle
\begin{abstract}
\nin A novel {\em Lattice Boltzmann Method} applicable to compressible fluid flows is developed. This method is based on replacing the governing equations by a relaxation system and the interpretation of the diagonal form of the relaxation system as a discrete velocity Boltzmann system. As a result of this interpretation, the local equilibrium distribution functions are simple algebraic functions of the conserved variables and the fluxes, without the low Mach number expansion present in the equilibrium distribution of the traditional Lattice Boltzmann Method (LBM). This new Lattice Boltzmann Relaxation Scheme (LBRS) thus overcomes the low Mach number limitation and can successfully simulate compressible flows.  While doing so, our algorithm retains all the distinctive features of the traditional LBM.  Numerical simulations carried out for inviscid flows in one and two dimensions show that the method can simulate the features of compressible flows like shock waves and expansion waves. 
\end{abstract}

\section{Introduction}
The lattice Boltzmann method (LBM) has emerged as an alternative to traditional CFD methods in the recent years. Its simplicity, ability to simulate complex fluid flow phenomena and amenability to parallel programming are among the reasons for its rising popularity. The lattice Boltzmann method is based on microscopic models and mesoscopic kinetic equations~\cite{Chen_Doolen}. The kinetic equations are simplified in the sense that the velocity space is discrete instead of the continuous velocity space used in kinetic theory. The lattice Boltzmann model thus consists of a small number of fictitious particles streaming (along certain fixed directions) and colliding on a lattice.  From this simplistic model, the macroscopic equations of fluid flows can be recovered by Chapman-Enskog expansion and thus such a simplistic model can be used to simulate fluid flows.  This connection between the mesoscopic and macroscopic levels allows LBM to simulate continuum fluid flows.  A key feature in favour of LBM is that the convection term of the lattice Boltzmann equation is linear as opposed to the nonlinear convection term in the macroscopic equations. Extensive reviews of the lattice Boltzmann method can be found in the papers of Chen and Doolen~\cite{Chen_Doolen} as well as that of Benzi {\em et al.}~\cite{Benzi_Succi_Vergasola} and in the books of Rothman \& Zaleski \cite{Rothman_Zaleski}, Wolf-Gladrow \cite{Wolf-Gladrow}, Rivet \& Boon \cite{Rivet_Boon}, Succi \cite{Succi}, Sukop \& Thorne \cite{Sukop_Thorne} and Mohamad \cite{Mohamad}. \\  

Historically, the lattice Boltzmann method was developed as an advancement over the lattice gas automata method~\cite{McNamara, Higuera, Chen_Chen}. Later, He and Luo~\cite{He_Luo} showed that the lattice Boltzmann equation can be derived directly from the Boltzmann equation. Irrespective of its numerous advantages, the lattice Boltzmann method has a serious drawback in that it is restricted to simulation of low Mach number (essentially incompressible) flows. There have been many attempts to extend this method to simulate compressible fluid flows. Alexander {\em et al.}~\cite{Alexander} have attempted to simulate Burgers equation with shocks using LBM. They have modified a parameter in the equilibrium distribution so that sound speed can be chosen arbitrarily. A major limitation of their model is that it can only simulate an isothermal fluid~\cite{Alexander}. Shouxin {\em et al.}~\cite{Shouxin} have used a 13-bit model based on the FHP lattice with the particles moving along the links being divided into two types,  each type having its own energy level and the rest particle having a different (third) energy level.  The equilibrium distribution is required to satisfy flux conditions besides the conservation of mass, momentum and energy. They have simulated the shock tube problem.  Guangwu {\em et al.}~\cite{Guangwu} have used a similar approach on a square lattice and have tested their model on three compressible flow test cases.  The drawback faced by both the models is that too many parameters have to be chosen. Sun~\cite{Sun} introduced a semi-discrete LB model wherein the particle velocities are determined by the mean flow velocity and internal energy. The particle velocities are adaptive which permits the mean flow to have a higher Mach number. The drawback of this model, as reported by Sun, is that the relaxation parameter has to be set equal to one. Kataoka and Tsutahara~\cite{Kataoka, Tsutahara} developed a lattice Boltzmann model for the compressible Euler and Navier-Stokes equations wherein the specific heat ratio can be chosen freely. Their model can simulate compressible fluid flows well but is unstable for Mach numbers greater than unity.  Tolke~\cite{Tolke} presented a new model based on the lattice Boltzmann method for simulation of thermal compressible flows. He used the lattice Boltzmann equation to solve for the flow field while the temperature equation is solved by a finite difference scheme. The sound speed varies proportionally with temperature to allow large density variations. The drawback of this model is that it is restricted to low Mach numbers while the thermal flows with large density variations can be simulated.  Yan {\em et al.}~\cite{Yan} proposed a Lagrangian LBM to simulate compressible isentropic flows. They have used displacement distribution functions instead of velocity distribution function.  While they have successfully simulated compressible inviscid flows, the authors report that issues regarding accuracy, stability and wall boundary condition for their method need to be solved. Zhang {\em et al.}~\cite{Zhang} and Yan {\em et al.}~\cite{Liu} have proposed a three-energy level and three-speed Lattice Boltzmann model using higher moments of the equilibrium distribution to simulate compressible flows.  Other authors~\cite{So,Erdembilegt,Fu} have used finite difference LBM with a modified equilibrium distribution function to overcome the low Mach number restriction of the traditional LB method.  Qu {\em et al.}~\cite{Qu} have developed an alternative method to construct the equilibrium distribution function to overcome the low Mach number restriction of the traditional LBM.\\

Although the list presented above is not exhaustive, it can be seen that attempts have been made in several directions to develop a robust LB method for compressible flows and yet, a widely-accepted route to tackle compressible flows using the lattice Boltzmann framework has not yet emerged.  In this paper, we have attempted to construct a compressible lattice Boltzmann method using a hitherto unexplored area, namely the framework of relaxation systems for hyperbolic conservation laws.  This Lattice Boltzmann relaxation scheme (LBRS) is based on the interpretation of the diagonal form of the relaxation system as a Lattice Boltzmann equation.  As a result, the local equilibrium distribution in our method is an algebraic function of the conserved variables and the fluxes. LBRS simulates compressible flows successfully as the local equilibrium distribution function is not based on the low Mach number expansion. The following section formulates the lattice Boltzmann relaxation scheme. In section III, compressible flow simulations for one and two dimensions are carried out using LBRS.  Section IV concludes this paper.

\section{Lattice Boltzmann Relaxation Scheme (LBRS)} 
The lattice Boltzmann relaxation scheme is based on the framework of a relaxation system. In this section, we first utilise the relaxation system of Jin \& Xin \cite{Jin_Xin} for a hyperbolic conservation law in one dimension to formulate our method. This relaxation system is then diagonalized and further interpreted as a discrete velocity Boltzmann equation, following Natalini \cite{Natalini}.  Then, we extend LBRS to two-dimensions.  The relaxation system of Jin and Xin cannot be diagonalized for the case of two-dimensions.  We use a novel diagonal relaxation system which is isotropic to extend our scheme to two-dimensions.  

\subsection{LBRS for scalar equations}
The relaxation system was introduced by Jin and Xin~\cite{Jin_Xin} to replace a nonlinear conservation law with a semi-linear hyperbolic system. We demonstrate the interpretation of this relaxation system as a lattice Boltzmann equation first for a one-dimensional hyperbolic conservation equation like the inviscid Burgers equation.

Consider the inviscid Burgers equation. 
\begin{equation} \label{1D_Burgers_Equation}
\frac{\partial{u}}{\partial{t}}+\frac{\partial{g(u)}}{\partial{x}} = 0,\qquad(x,t)\>\in\>R^1\times{R_+},\qquad u\>\in\>R^1
\end{equation}

with initial data $ u(x,0) = u_0(x) $ and $g(u)= \frac{1}{2} u^2$

The relaxation system introduced by Jin and Xin~\cite{Jin_Xin} is given as:
\begin{align}
\begin{split}
\frac{\partial{u}}{\partial{t}}+\frac{\partial{v}}{\partial{x}} &= 0,\qquad{v}\>\in\>R^1 \\
\frac{\partial{v}}{\partial{t}}+\lambda^2\frac{\partial{u}}{\partial{x}} &=-\frac{1}{\epsilon}(v-g(u))
\end{split}
\end{align}

where $\epsilon$ is the relaxation parameter and $\lambda$ is a positive constant to be determined from the sub-characteristic condition.  Note that as $\epsilon \rightarrow 0$, the second equation of the relaxation system gives $v = g(u)$ which, when substituted into the first equation of the relaxation system leads to the original conservation equation (\ref{1D_Burgers_Equation}).  The relaxation system can be written in vector form as follows:
\begin{equation}\label{vector_form}
\frac{\partial{\mathbf{Q}}}{\partial{t}}+\mathbf{A}\frac{\partial\mathbf{Q}}{\partial{x}} = \mathbf{H}
\end{equation}
where\qquad $\textbf{Q}\>=\left[{\begin{array}{cc}
                   u\\
                   v\\
                  \end{array}}\right] \> ;\qquad  \textbf{A}\>=\left[{\begin{array}{cc}
                   0 & 1\\
                   \lambda^2 & 0\\
                  \end{array}}\right] \> ;\qquad \textbf{H}\>=\left[{\begin{array}{cc}
                   0\\
                   -\frac{1}{\epsilon}(v-g(u))\\
                  \end{array}}\right] $\\

Note that the eigenvalues of $A$ are real and distinct ($\pm\lambda$) and thus the relaxation system is strictly hyperbolic.  Therefore, the diagonal form of the relaxation system is obtained by introducing characteristic variable vector 
 $\mathbf{f}=R^{-1}\mathbf{Q}$ in (\ref{vector_form}) where $R$ is the right eigenvector of $A$, given by the following expression.  
$$ R = \left[ \begin{array}{cc} 1 & 1 \\ -\lambda & \lambda \end{array}\right] \ \textrm{and} \ R^{-1} = \left[ \begin{array}{cc} \fr{1}{2} & - \fr{1}{2 \lambda} \\[2mm] \fr{1}{2} & \fr{1}{2 \lambda} \end{array} \right]$$   
Thus, we obtain   
\begin{equation}\label{diagonal_form}
\frac{\partial{\mathbf{f}}}{\partial{t}}+\mathbf{\Lambda}\frac{\partial\mathbf{f}}{\partial{x}} = R^{-1}\mathbf{H}
\end{equation}
where 
$$\textbf{f}\>=\left[{\begin{array}{c}
                   f_1\\
                   f_2\\
                  \end{array}}\right]=\left[{\begin{array}{c}
                   \frac{u}{2}-\frac{v}{2\lambda}\\
                   \frac{u}{2}+\frac{v}{2\lambda}\\
                  \end{array}}\right] \> ;\quad \mathbf{\Lambda}=R^{-1}\mathbf{A}R=\left[{\begin{array}{cc}
                   -\lambda & 0\\
                   0 & \lambda\\
                  \end{array}}\right] \> ; \quad  R^{-1}\mathbf{H}=\left[{\begin{array}{c}
                   \frac{1}{2\lambda\epsilon}[v-g(u)]\\
                   -\frac{1}{2\lambda\epsilon}[v-g(u)]\\
                  \end{array}}\right] $$

The vector form can be written in terms of the components as
\begin{align}\label{equation_for_f}
\begin{split}
\frac{\partial{f_1}}{\partial{t}}-\lambda\frac{\partial{f_1}}{\partial{x}} &= \frac{1}{2\lambda\epsilon}[v-g(u)] \\
\frac{\partial{f_2}}{\partial{t}}+\lambda\frac{\partial{f_2}}{\partial{x}} &=-\frac{1}{2\lambda\epsilon}[v-g(u)]
\end{split}
\end{align}

This system can be interpreted as a discrete velocity Boltzmann equation if we introduce the following new variables:
\begin{equation}\label{equilibrium}
f^{eq}_1=\frac{u}{2}-\frac{g(u)}{2\lambda}\qquad and \qquad f^{eq}_2=\frac{u}{2}+\frac{g(u)}{2\lambda}
\end{equation}

\begin{figure}[H]
\includegraphics[width=5cm]{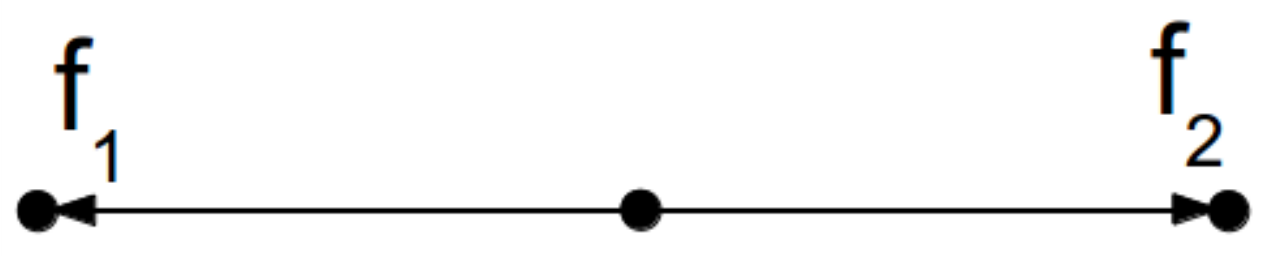}
\centering
\caption{D1Q2 model}
\label{fig:D1Q2}
\end{figure}

Equation \ref{equation_for_f} now transforms to
\begin{align}\label{LBRS_equation}
\begin{split}
\frac{\partial{f_1}}{\partial{t}}-\lambda\frac{\partial{f_1}}{\partial{x}} &=-\frac{1}{\epsilon}[f_1-f^{eq}_1]\\
\frac{\partial{f_2}}{\partial{t}}+\lambda\frac{\partial{f_2}}{\partial{x}} &=-\frac{1}{\epsilon}[f_1-f^{eq}_2]
\end{split}
\end{align}

with $f_1$ and $f_2$ given by equation \ref{diagonal_form}.  Thus, $-\lambda$ \& $\lambda$ are the discrete velocities and $f_{1}$ \& $f_{2}$ are the corresponding components of the distribution function $\mathbf{f}$ in the discrete velocity Boltzmann equation with the B-G-K model given by 
\begin{equation} \label{1D_DVBE}  
\fr{\partial \mathbf{f}}{\partial t} + \Lambda \fr{\partial \mathbf{f}}{\partial x} = - \fr{1}{\epsilon} \left[ \mathbf{f} - \mathbf{f}^{eq} \right] 
\end{equation} 
This interpretation was given by Natalini \cite{Natalini}.  
The initial condition $u(x,0)=u_0(x)$ can be rewritten as $f_i=f^{eq}_i(u_0(x))$.
Equations (\ref{LBRS_equation}) resemble a two-velocity lattice BGK model with the distribution function $\mathbf{f}$ relaxing to the equilibrium distribution function $\mathbf{f}^{eq}$ within the relaxation time $\epsilon$. This interpretation of the relaxation system is the basis of the Lattice Boltzmann Relaxation Scheme (LBRS). In accordance with the terminology adopted for lattices in the standard LB literature, this model can be termed as the D1Q2 model with two particles traveling in opposite directions with speed $\lambda$ at every lattice node as shown in figure (\ref{fig:D1Q2}). It is to be noted that the equilibrium distribution functions in LBRS defined by equation (\ref{equilibrium}) are simple algebraic functions of the conserved variable and the associated flux. These functions are not expressed as a polynomial (Gaussian) function of the macroscopic velocity (as is the case in the traditional lattice Boltzmann method) and hence our method is able to simulate compressible flows as no low Mach number expression is involved. The left hand side of the equation represents convection on the lattice  while the right hand side represents collision at a lattice node.  The LBRS algorithm retains the hopping-collision cycle of the traditional method.  The variables $u,v$ and $g(u)$ of the relaxation system are recovered as a simple summation of the distribution functions as illustrated in equation (\ref{moments}). The original conservation law is then recovered from the relaxation system as the relaxation time tends to zero. Thus our method successfully overcomes the low Mach number restriction plaguing the traditional LB method while maintaining the structure and ease of programming of the LBM.    

\begin{equation}\label{moments}
u=f_1+f_2=f^{eq}_1+f^{eq}_2 \>;\qquad v=-\lambda f_1+ \lambda f_2 \>; \qquad g(u)=-\lambda f^{eq}_1+ \lambda f^{eq}_2
\end{equation}

Another feature of the traditional LB method that carries over to LBRS is that the number of particles in the model can be increased as long as the moment relations are satisfied. Generalizing to $n$ particles, the moment relations are
\begin{equation}\label{moments_n-particles}
u=\sum\limits_{i=1}^nf_i=\sum\limits_{i=1}^nf^{eq}_i \>;\qquad v=\sum\limits_{i=1}^n\lambda_if_i \>; \qquad g(u)=\sum\limits_{i=1}^n\lambda_if^{eq}_i
\end{equation}

The speed of the lattice particle ($\lambda$) is determined from the sub-characteristic condition defined in Jin and Xin~\cite{Jin_Xin}:
\begin{equation}
\lambda^2\geq\Bigg(\frac{\partial{g(u)}}{\partial{u}}\Bigg)^2 \qquad or \qquad -\lambda\leq\frac{\partial{g(u)}}{\partial{u}}\leq\lambda
\end{equation}

The equation (\ref{1D_DVBE}) can be written in discrete from as follows:
\begin{equation}\label{discrete_LBRS}
f_i(x+\Delta{x},t+\Delta{t})=f_i(x,t)-\frac{\Delta{t}}{\epsilon}[f_i(x,t)-f^{eq}_i(x,t)] \>; \qquad for \> i=1,...,n
\end{equation}

The time interval $\Delta{t}$ is chosen such that convection is exact {\em i.e.}, in the time interval $\Delta{t}$, the particles hop exactly from one lattice node to the adjacent node in the corresponding direction. The discrete form (equation (\ref{discrete_LBRS})) can thus be written as
\begin{equation}
f_i(x+\lambda_i\Delta{t},t+\Delta{t})=(1-\omega)f_i(x,t)+\omega f^{eq}_i(x,t)] \>; \qquad for \> i=1,...,n
\end{equation}

where $\omega=\Delta{t}/\epsilon$ and takes a value between 0 and 2. The limits for the value of $\omega$ are decided based on a Chapman-Enskog type analysis of the LBRS equations.  We present the analysis for the case of one-dimensional Burgers equation in the following subsection.

\subsection{Chapman-Enskog analysis of LBRS}
The macroscopic equations are derived from the traditional lattice Boltzmann equations by carrying out a multi-scale analysis which is similar to the Chapman-Enskog analysis in deriving Navier-Stokes equations from the classical Boltzmann equation.  As a result of this Chapman-Enskog type analysis, the diffusive term and the corresponding transport coefficient can be recovered.  Since the relaxation approximation is known to be a vanishing diffusion model to the original governing equation, we have carried out a similar multi-scale analysis for our scheme, for the case of one-dimensional Burgers equation, to determine the diffusion coefficient. The limits for the relaxation parameter are then decided based on the physically apt values of the diffusion coefficient.  

Consider the discrete LBRS equation given by equation (\ref{discrete_LBRS}).  
\begin{equation}
f_i(x+\Delta{x},t+\Delta{t})=f_i(x,t)-\frac{\Delta{t}}{\epsilon}[f_i(x,t)-f^{eq}_i(x,t)] \>; \qquad for \> i=1,...,n
\end{equation}

Substituting Taylor series expansion for the term on the LHS of this equation and imposing the exact hopping condition ($\lambda_i \Delta{t} = \Delta{x}$) results in\\
\begin{equation} 
\lambda_i \Delta{t} \> \frac{\partial{f_i}}{\partial{x}} + \Delta{t} \> \frac{\partial{f_i}}{\partial{t}} + \frac{\lambda^{2}_i \Delta{t^2}}{2} \> \frac{\partial^2{f_i}}{\partial{x^2}} + \frac{\Delta{t^2}}{2} \> \frac{\partial^2{f_i}}{\partial{t^2}} +\lambda_i\Delta{t^2} \> \frac{\partial^2{f_i}}{\partial{x}\partial{t}}+\omega f_i - \omega f^{eq}_i = 0
\end{equation}

Introducing the time and space scales corresponding to relaxation ($t_r,\> x$), convection ($t_1 = \varepsilon^{-1} t_r, \> x_1 = \varepsilon^{-1} x$) and diffusion ($t_2 = \varepsilon^{-2} t_r,\> x_2 = \varepsilon^{-1} x$) phenomena, the temporal and spatial derivatives are given by\\
\begin{equation}
\frac{\partial}{\partial{t}} = \varepsilon \frac{\partial}{\partial{t_1}} + \varepsilon^2 \frac{\partial}{\partial{t_2}} \>;  \qquad \qquad \frac{\partial}{\partial{x}} = \varepsilon \frac{\partial}{\partial{x_1}}
\end{equation}

\begin{equation}
\frac{\partial^2}{\partial{t^2}} = \varepsilon^2 \frac{\partial^2}{\partial{{t^2}_1}} + 2 \varepsilon^3 \frac{\partial^2}{\partial{t_1}\partial{t_2}} + \varepsilon^4 \frac{\partial^2}{\partial{{t^2}_2}}
\end{equation}

\begin{equation}
\frac{\partial^2}{\partial{t}\partial{x}} = \varepsilon^2 \frac{\partial^2}{\partial{t_1}\partial{x_1}} + \varepsilon^3 \frac{\partial^2}{\partial{t_2}\partial{x_1}} \>; \qquad \qquad \frac{\partial^2}{\partial{x^2}} = \varepsilon^2 \frac{\partial^2}{\partial{{x^2}_1}}
\end{equation}

Based on the same scalings, we expand $f_i$ as 
\begin{equation} \label{expansion_of_f}
f_i = f^{eq}_i + \varepsilon f^{(1)}_i + \varepsilon^2 f^{(2)}_i + O(\varepsilon^3)
\end{equation} 

Upon substituting the scaled time and space derivatives as well as the expansion of the distribution function, we obtain the following expressions

\begin{equation}
O(\varepsilon): \qquad \frac{\partial{f^{eq}_i}}{\partial{t_1}} + \lambda_i \frac{\partial{f^{eq}_i}}{\partial{x_1}} + \frac{\omega}{\Delta{t}} f^{(1)}_i = 0
\end{equation}

\begin{equation}
O(\varepsilon^2): \qquad  \frac{\partial{f^{(1)}_i}}{\partial{t_1}} + \frac{\partial{f^{eq}_i}}{\partial{t_2}} + \lambda_i \frac{\partial{f^{(1)}_i}}{\partial{x_1}} + \frac{\lambda^2_i \Delta{t}}{2}\frac{\partial^2{f^{eq}_i}}{\partial{x{^2}_1}} + \lambda_i \Delta{t} \frac{\partial^2{f^{eq}_i}}{\partial{x_1}\partial{t_1}} + \frac{\Delta{t}}{2} \frac{\partial^2{f^{eq}_i}}{\partial{t{^2}_1}} + \frac{\omega}{\Delta{t}} f^{(2)}_i = 0
\end{equation}

We have separated the $O(\varepsilon)$ and the $O(\varepsilon^2)$ terms and hence we obtain two expressions. We now take moments of these terms based on the moment relations given by equations (\ref{moments_n-particles}).  Noting that the moments of both the distribution function and its equilibrium part must be the same, the zeroth moment of the $O(\varepsilon)$ terms results in the following expression
\begin{equation}\label{moment_of_epsilon}
\varepsilon \Bigg(\frac{\partial{u}}{\partial{t_1}} + \frac{\partial{g(u)}}{\partial{x_1}}\Bigg)= 0
\end{equation}

The zeroth moment of the $O(\varepsilon^2)$ terms gives
\begin{equation}\label{moment_of_epsilonsquared}
\varepsilon^2 \frac{\partial{u}}{\partial{t_2}} =\varepsilon^2 \Delta{t} \Bigg(\frac{1}{\omega}-\frac{1}{2}\Bigg) \Bigg[ \frac{\partial^2{\sum\limits_{i=1}^n\lambda^2_i f^{eq}_i}}{\partial{x{^2}_1}} - \frac{\partial^2{\frac{1}{3}u^3}}{\partial{x{^2}_1}}\Bigg]
\end{equation}

Upon addition of equations (\ref{moment_of_epsilon}) and (\ref{moment_of_epsilonsquared}) and using the definitions for the  scaled time and space derivatives we recover the original macroscopic equation with a diffusive correction as shown below.
\begin{equation}
\frac{\partial{u}}{\partial{t}} + \frac{\partial{g(u)}}{\partial{x}} = \Delta{t} \Bigg(\frac{1}{\omega}-\frac{1}{2}\Bigg) \Bigg[ \frac{\partial^2{\sum\limits_{i=1}^n\lambda^2_i f^{eq}_i}}{\partial{x{^2}}} - \frac{\partial^2{\frac{1}{3}u^3}}{\partial{x{^2}}}\Bigg]
\end{equation}

Here $g(u) = \frac{1}{2} u^2$.  We can conclude from the previous equation that $\omega$ must take values in the interval (0, 2) to ensure a positive viscosity coefficient. 

\subsection{LBRS for one-dimensional system of equations}
Now that we have discussed our method in detail for the one-dimensional scalar equations, we extend it to systems of conservation laws in one and two dimensions. In this subsection, we demonstrate LBRS for 1-D Euler equations which are given by
\begin{equation}
\frac{\partial{\mathbf{U}}}{\partial{t}}+\frac{\partial{\mathbf{G(U)}}}{\partial{x}}=0
\end{equation}\\
where 
$$\textbf{U}\>=\left[{\begin{array}{cc}
                   \rho\\
                   \rho u\\
                   \rho E\\
                  \end{array}}\right]$$ 
is the vector of conserved variables and 
$$\mathbf{G(U)}=\left[{\begin{array}{cc}
                   \rho u\\
                   p+\rho u^2\\
                   pu+\rho uE\\
                  \end{array}}\right]$$ 
is the flux vector. $E=\fr{p}{\rho(\gamma-1)}+\fr{1}{2}u^2$ is the total energy and the equation of state is given by $p=\rho RT$. The equations for the lattice Boltzmann relaxation scheme for this case are similar to equations (\ref{LBRS_equation}) except that the distribution functions are now vectors themselves. This is because each of the mass, momentum and energy conservation equations is now represented by a 2-velocity diagonal relaxation system. The LBRS equations for the D1Q2 model for the one-dimensional Euler system are
\begin{align}
\begin{split}
\frac{\partial{\mathbf{f_1}}}{\partial{t}}-\lambda\frac{\partial{\mathbf{f_1}}}{\partial{x}} &= -\frac{1}{\epsilon}[\mathbf{f_1}-\mathbf{f^{eq}_1}]\\
\frac{\partial{\mathbf{f_2}}}{\partial{t}}+\lambda\frac{\partial{\mathbf{f_2}}}{\partial{x}} &= -\frac{1}{\epsilon}[\mathbf{f_2}-\mathbf{f^{eq}_2}]
\end{split}
\end{align}\\

where \; $\mathbf{f_1}\>=\left[{\begin{array}{c}
                   f_{11}\\
                   f_{12}\\
                   f_{13}\\
                  \end{array}}\right] \> ; \quad \mathbf{f_2}=\left[{\begin{array}{c}
                   f_{21}\\
                   f_{22}\\
                   f_{23}\\
                  \end{array}}\right] \> ; \qquad \mathbf{f^{eq}_1}=\left[{\begin{array}{c}
                   f^{eq}_{11}\\
                   f^{eq}_{12}\\
                   f^{eq}_{13}\\
                  \end{array}}\right] \> ; \qquad \mathbf{f^{eq}_2}=\left[{\begin{array}{c}
                   f^{eq}_{21}\\
                   f^{eq}_{22}\\
                   f^{eq}_{23}\\
                  \end{array}}\right]$\\ 

The moment relations are given by

\begin{equation}\label{moments_system}
\mathbf{U}\>=\left[{\begin{array}{cc}
                   \rho\\
                   \rho u\\
                   \rho E\\
                  \end{array}}\right]=\sum\limits_{i=1}^n\left[{\begin{array}{cc}
                   f_{i1}\\
                   f_{i2}\\
                   f_{i3}\\
                  \end{array}}\right]=\sum\limits_{i=1}^n\left[{\begin{array}{cc}
                   f^{eq}_{i1}\\
                   f^{eq}_{i2}\\
                   f^{eq}_{i3}\\
                  \end{array}}\right] \> ; \quad \mathbf{G(U)}\>=\left[{\begin{array}{cc}
                   \rho u\\
                   P+\rho u^2\\
                   Pu+\rho uE\\
                  \end{array}}\right]=\sum\limits_{i=1}^n \lambda_i \left[{\begin{array}{cc}
                   f^{eq}_{i1}\\
                   f^{eq}_{i2}\\
                   f^{eq}_{i3}\\
                  \end{array}}\right]
\end{equation}
where $n$ is the number of particles. The expressions for the equilibrium distribution functions are
\begin{equation}
f^{eq}_{1,k}=\frac{\mathbf{U}_{k}}{2}-\frac{\mathbf{G(U)_{k}}}{2\lambda} \>; \quad f^{eq}_{2,k}=\frac{\mathbf{U}_{k}}{2}+\frac{\mathbf{G(U)_{k}}}{2\lambda} \>, \quad k=1,2,3
\end{equation}

The speed $\lambda$ of the particles is calculated from the sub-characteristic condition specified by Jin and Xin~\cite{Jin_Xin} for systems of conservation laws. Out of the two choices proposed by them, we use the following one:
\begin{equation}
\lambda=max\big[|u-a|,|u|,|u+a|\big]
\end{equation}

\begin{figure}[H]
\includegraphics[width=5cm]{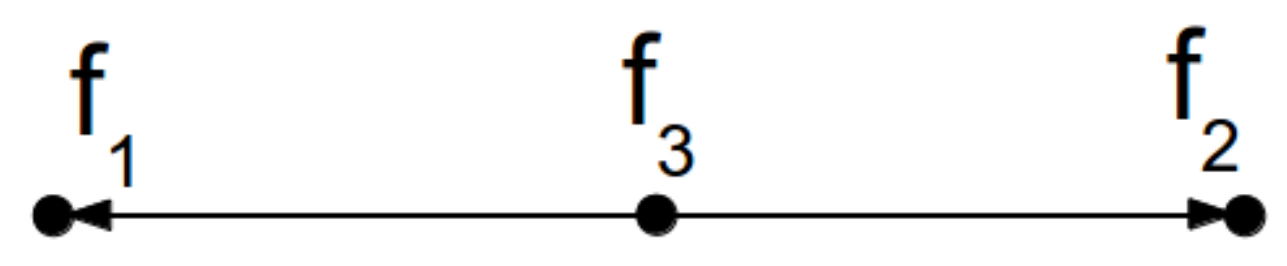}
\centering
\caption{D1Q3 model}
\label{fig:D1Q3}
\end{figure}

We demonstrate in the next section that the D1Q2 (2-velocity) model, though it simulates the features of 1-D Euler test cases, needs some improvement.  Thus we introduce here the D1Q3 model by adding a rest particle to the 2-velocity model.  We then test this model for the one-dimensional Euler case.  The D1Q3 model, (figure \ref{fig:D1Q3}) which is the standard model for 1-D flows in the traditional LB method, overcomes the flaws exhibited by the D1Q2 model as will be elaborated in the next section.    

The equations for the D1Q3 model are:
\begin{align}
\begin{split}
\frac{\partial{\mathbf{f_1}}}{\partial{t}}-\lambda\frac{\partial{\mathbf{f_1}}}{\partial{x}} &= -\frac{1}{\epsilon}[\mathbf{f_1}-\mathbf{f^{eq}_1}]\\
\frac{\partial{\mathbf{f_2}}}{\partial{t}} &= -\frac{1}{\epsilon}[\mathbf{f_2}-\mathbf{f^{eq}_2}]\\
\frac{\partial{\mathbf{f_3}}}{\partial{t}}+\lambda\frac{\partial{\mathbf{f_3}}}{\partial{x}} &= -\frac{1}{\epsilon}[\mathbf{f_3}-\mathbf{f^{eq}_3}]
\end{split}
\end{align}

For simplicity, we write the above equations in diagonal form:
\begin{equation}
\frac{\partial{\mathbf{f}}}{\partial{t}}+\mathbf{\Lambda}\frac{\partial\mathbf{f}}{\partial{x}}=-\frac{1}{\epsilon}[\mathbf{f}-\mathbf{f^{eq}}]
\end{equation}

where $\mathbf{f}$ and $\mathbf{f^{eq}}$ are 3-dimensional vectors and $\mathbf{\Lambda}$ is a $3\times3$ diagonal matrix given by\\

\qquad \qquad $\mathbf{f}\>=\left[{\begin{array}{ccc}
                   \mathbf{f_1}\\
                   \mathbf{f_2}\\
                   \mathbf{f_3}\\
                  \end{array}}\right] \> ; \qquad \mathbf{\Lambda}=\left[{\begin{array}{ccc}
                   -\lambda & 0 & 0\\
                   0 & 0 & 0\\
                   0 & 0 & \lambda\\
                  \end{array}}\right] \> ; \qquad  \mathbf{f^{eq}}=\left[{\begin{array}{ccc}
                   \mathbf{f^{eq}_1}\\
                   \mathbf{f^{eq}_2}\\
                   \mathbf{f^{eq}_3}\\
                  \end{array}}\right]$\\

The moment relations for this model remain unchanged from equations (\ref{moments_system}) except that $n$ (number of particles) is now equal to 3. The unchanged moment relations allow us to recover the governing equations (in the zero relaxation limit) irrespective of the number of particles used in the model. Addition of an extra particle however alters the equilibrium distribution functions. The equilibrium distribution functions for the D1Q3 model are evaluated from the following moment relations:
\begin{equation}\label{D1Q3_moments}
\qquad\mathbf{U}_{k}=\sum\limits_{i=1}^3f_{i,k}=\sum\limits_{i=1}^3f^{eq}_{i,k} \>; \qquad \qquad \mathbf{G(U)}_{k}=\sum\limits_{i=1}^3\lambda_i\>f^{eq}_{i,k}\>,\qquad \qquad k=1,2,3   \qquad \qquad
\end{equation}

Assuming the equilibrium distribution functions to be a linear combination of $\mathbf{U}_{k}$ and $\mathbf{G(U)_{k}}$ we obtain expressions for $f^{eq}_{i,k}, i,k=1,2,3$, subject to conditions (\ref{D1Q3_moments}) 
\begin{equation}
f^{eq}_{1,k}=\frac{\mathbf{U}_{k}}{3}-\frac{\mathbf{G(U)}_{k}}{2\lambda} \>; \qquad f^{eq}_{2,k}=\frac{\mathbf{U}_{k}}{3} \>; \qquad f^{eq}_{3,k}=\frac{\mathbf{U}_{k}}{3}+\frac{\mathbf{G(U)}_{k}}{2\lambda}, \qquad k=1,2,3 \quad
\end{equation}

We now move on to demonstrating the extension of LBRS to two dimensions.

\subsection{LBRS for two-dimensions}
\begin{figure}[H]
\includegraphics[width=5cm]{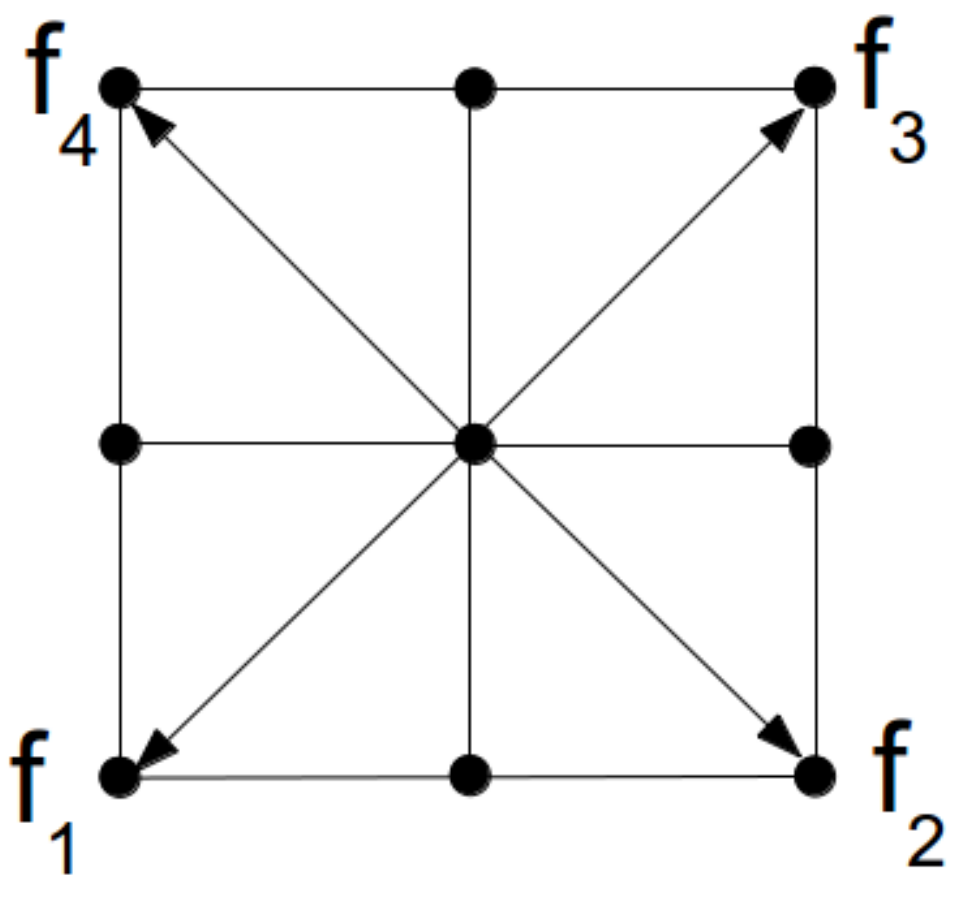}
\centering
\caption{D2Q4 (Isotropic relaxation system) model}
\label{fig:D2Q4}
\end{figure}

As an improvement over the relaxation system introduced by Jin and Xin \cite{Jin_Xin} for two-dimensions, Aregba-Driollet \& Natalini \cite{Driollet_Natalini} introduced diagonal discrete velocity systems in 2-D.  However, these systems are not isotropic and we use the isotropic (diagonal) relaxation system introduced by Raghurama Rao and used by Jayaraj \cite{Jayaraj} and Arun {\em et al.}~\cite{Arun_RaghuramaRao_1, Arun_RaghuramaRao_2}. This is a D2Q4 model with four particles travelling in the diagonal directions with speeds equal to $\sqrt{2}\>\lambda$ at each node (see figure (\ref{fig:D2Q4})). Let us consider a two-dimensional scalar conservation law:
\begin{equation}\label{conservation_law}
\frac{\partial{u}}{\partial{t}}+\frac{\partial{g_1(u)}}{\partial{x}}+\frac{\partial{g_2(u)}}{\partial{x}}=0
\end{equation}

The equations for the D2Q4 model which is a relaxation approximation for this scalar conservation law are as follows:
\begin{align}\label{D2Q4equations}
\begin{split}
\frac{\partial{f_1}}{\partial{t}}-\lambda\frac{\partial{f_1}}{\partial{x}}-\lambda\frac{\partial{f_1}}{\partial{y}} &= -\frac{1}{\epsilon}[f_1-f^{eq}_1]\\
\frac{\partial{f_2}}{\partial{t}}+\lambda\frac{\partial{f_2}}{\partial{x}}-\lambda\frac{\partial{f_2}}{\partial{y}} &= -\frac{1}{\epsilon}[f_2-f^{eq}_2]\\
\frac{\partial{f_3}}{\partial{t}}+\lambda\frac{\partial{f_3}}{\partial{x}}+\lambda\frac{\partial{f_3}}{\partial{y}} &= -\frac{1}{\epsilon}[f_3-f^{eq}_3]\\
\frac{\partial{f_4}}{\partial{t}}-\lambda\frac{\partial{f_4}}{\partial{x}}+\lambda\frac{\partial{f_4}}{\partial{y}} &= -\frac{1}{\epsilon}[f_4-f^{eq}_4]
\end{split}
\end{align}

Equations \ref{D2Q4equations} can be written in vector form as
\begin{equation}
\frac{\partial{\mathbf{f}}}{\partial{t}}+\mathbf{\Lambda_1}\frac{\partial\mathbf{f}}{\partial{x}}+\mathbf{\Lambda_2}\frac{\partial\mathbf{f}}{\partial{y}}=-\frac{1}{\epsilon}[\mathbf{f}-\mathbf{f^{eq}}]
\end{equation}

with  \\ \\ $\mathbf{f}=\left[{\begin{array}{c}
                   f_1\\
                   f_2\\
                   f_3\\
                   f_4\\
                  \end{array}}\right] \> ; \; \mathbf{\Lambda_1}=\left[{\begin{array}{cccc}
                   -\lambda & 0 & 0 & 0\\
                   0 & \lambda & 0 & 0\\
                   0 & 0 & \lambda & 0\\
                   0 & 0 & 0 & -\lambda\\
                  \end{array}}\right] \> ; \; \mathbf{\Lambda_2}=\left[{\begin{array}{cccc}
                   -\lambda & 0 & 0 & 0\\
                   0 & -\lambda & 0 & 0\\
                   0 & 0 & \lambda & 0\\
                   0 & 0 & 0 & \lambda\\
                  \end{array}}\right] \> ; \; \mathbf{f^{eq}}=\left[{\begin{array}{c}
                   f^{eq}_1\\
                   f^{eq}_2\\
                   f^{eq}_3\\
                   f^{eq}_4\\
                  \end{array}}\right]$\\

The equilibrium distribution functions are determined from the moment conditions:
\begin{equation}\label{D2Q4moments}
\qquad u=P\mathbf{f}=P\mathbf{f^{eq}} \>; \qquad g_1(u)=P\mathbf{\Lambda_1}\mathbf{f^{eq}} \>; \qquad g_2(u)=P\mathbf{\Lambda_2f^{eq}} \qquad
\end{equation}

where $P=[1\;1\;1\;1]$, $u$ is the conserved variable, $g_1(u)$ and $g_2(u)$ are fluxes in the $x$ and $y$ directions respectively.

Again assuming the equilibrium distribution functions to be a linear combination of the conserved variable and the fluxes, we obtain expressions for the equilibrium distribution functions subject to conditions (\ref{D2Q4moments}).
\begin{equation}
\mathbf{f^{eq}}=\left[{\begin{array}{cc}
                   f^{eq}_1\\
                   f^{eq}_2\\
                   f^{eq}_3\\
                   f^{eq}_4\\
                  \end{array}}\right]=\left[{\begin{array}{cc}
                   \frac{u}{4}-\frac{g_1(u)}{4\lambda}-\frac{g_2(u)}{4\lambda}\\
                   \frac{u}{4}+\frac{g_1(u)}{4\lambda}-\frac{g_2(u)}{4\lambda}\\
                   \frac{u}{4}+\frac{g_1(u)}{4\lambda}+\frac{g_2(u)}{4\lambda}\\
                   \frac{u}{4}-\frac{g_1(u)}{4\lambda}+\frac{g_2(u)}{4\lambda}\\
                  \end{array}}\right]
\end{equation}

To recover the relaxation system corresponding to the D2Q4 model (the isotropic relaxation system), we multiply equations \ref{D2Q4equations} by $P$, $P\Lambda_1$ and $P\Lambda_2$ respectively to get
\begin{align}
\begin{split}
\frac{\partial{u}}{\partial{t}}+\frac{\partial{v_1}}{\partial{x}}+\frac{\partial{v_2}}{\partial{y}}&=0\\
\frac{\partial{v_1}}{\partial{t}}+\lambda^2\frac{\partial{u}}{\partial{x}}+\frac{\partial{w}}{\partial{y}}&=-\frac{1}{\epsilon}[v_1-g_1(u)]\\
\frac{\partial{v_2}}{\partial{t}}+\frac{\partial{w}}{\partial{x}}+\lambda^2\frac{\partial{u}}{\partial{y}}&=-\frac{1}{\epsilon}[v_2-g_2(u)]\\
\end{split}
\end{align}

where $v_1 = P\Lambda_1\mathbf{f}$, $v_2 = P\Lambda_2\mathbf{f}$ and $w=P\Lambda_1\Lambda_2\mathbf{f}=P\Lambda_2\Lambda_1\mathbf{f}=0$.  Thus, we recover the relaxation system of Jin and Xin \cite{Jin_Xin}.  In the limit $\epsilon \rightarrow 0$ the original conservation law (equation (\ref{conservation_law})) is recovered. \\ 

The diagonal relaxation system for a system of conservation laws is similar to equation (\ref{D2Q4equations}) except that the components of the distribution function vector are vectors themselves {\em i.e.}, each equation of the system is represented by a diagonal relaxation approximation given by equations (\ref{D2Q4equations}).  For a multi-dimensional system of hyperbolic conservation laws, the stability condition proposed by Bouchut~\cite{Bouchut} must be used. This condition states that the eigenspectrum of $\mathbf{M}^{'}(u)$ must be positive, {\em i.e.}, 
\begin{equation}\label{Bouchut's_condition}
\qquad \sigma\Big(\mathbf{M}^{'}(u)\Big)\subset{[0,+\infty[} \qquad
\end{equation}

If condition (\ref{Bouchut's_condition}) is satisfied then there exists a kinetic entropy for the relaxation approximation associated with any convex entropy  $\eta$ of the original conservation law and Lax entropy inequalities are satisfied in the limit $\epsilon \rightarrow 0 $. Condition (\ref{Bouchut's_condition}) for the D2Q4 model simulating the two-dimensional Euler equations gives 
\begin{equation}
\lambda = max\bigg(|u+v|,|u+v+\sqrt(2a)|,|u+v-\sqrt(2a)|,|u-v|,|u-v-\sqrt(2a)|,|u-v+\sqrt(2a)|\bigg)
\end{equation}

where $u,v$ are the macroscopic velocities in the $x,y$ directions respectively and $a$ is the speed of sound.\\

For a typical multi-dimensional discrete velocity Boltzmann equation of the type 
$$ \fr{\partial f_{i}}{\partial t} + \vec{\lambda} \cdot \fr{\partial f_{i}}{\partial \vec{r}} = - \fr{1}{\epsilon} \left[ f_{i} - f^{eq}_{i} \right]$$ 
the LBRS equation is given by 
$$ f_{i} \left( \vec{r}_{i} + \vec{\lambda}_{i} \Delta t, t+\Delta t \right) = \left( 1 - \omega \right) f_{i} \left( \vec{r}_{i}, t \right) + \omega f^{eq}_{i}  $$   
where $\omega = \fr{\Delta t}{\epsilon}$.  The solution, as in the 1-D case, consists of a streaming step and a collision (relaxation) step, with streaming in the appropriate directions dictated by the discrete velocity model (in our D2Q4 case, the diagonal directions).   
    
\subsection{Implementation of wall boundary conditions}
For the lattice Boltzmann relaxation scheme presented in the previous section, we need to add a strategy for implementation of the wall boundary conditions for our method.  We will discuss these boundary conditions for the inviscid two-dimensional Euler equations.  As the flow is inviscid, the appropriate wall boundary condition is the free-slip or flow tangency boundary condition.  In the traditional lattice Boltzmann method, the principle of specular reflection is used at the wall to enforce the free-slip boundary condition~\cite{Succi}.  We incorporate this principle in our framework. 

\begin{figure}[H]
\includegraphics[width=5cm]{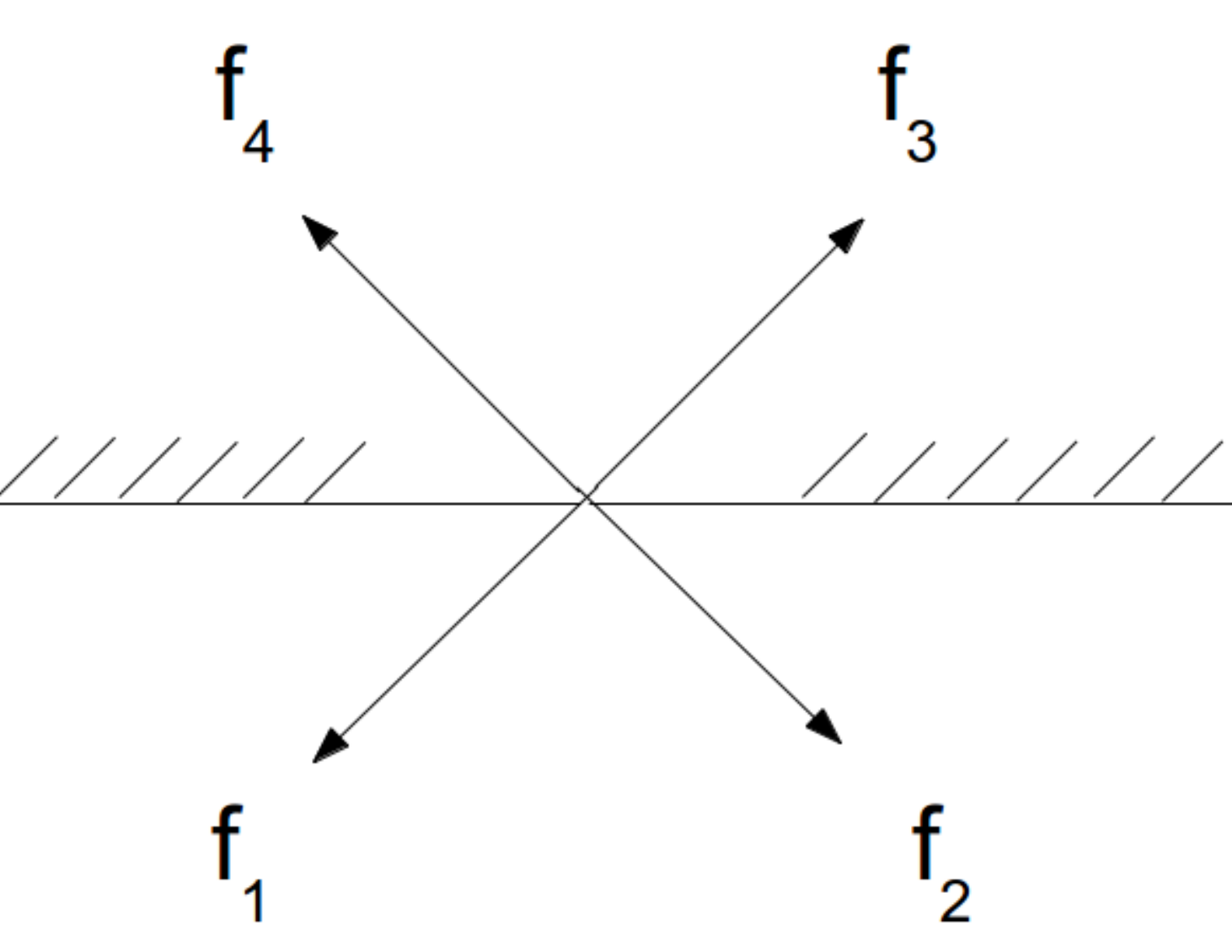}
\centering
\caption{D2Q4 model at an upper wall}
\label{fig:BCimplementation}
\end{figure}

The free-slip boundary condition implies that the normal component of the velocity at the wall is zero and that the flow is purely tangential at the wall.  In our framework, the normal component of velocity (actually the flux involving the normal velocity component) is expressed as a moment of the distribution functions. Let us consider an upper wall (figure (\ref{fig:BCimplementation})) where the free-slip condition is to be imposed.

For the continuity equation we have, 
$\rho v= P\mathbf{\Lambda_2} \mathbf{f^a}=\lambda(-f^a_1-f^a_2+f^a_3+f^a_4)=0$

This condition is satisfied when
\begin{equation}\label{f1_equation}
f^a_1=f^a_4
\end{equation}
\begin{equation}\label{f2_equation}
f^a_2=f^a_3
\end{equation}

The superscript $a$ is used to identify the distribution functions for the continuity equation. As a result of equations \ref{f1_equation} and \ref{f2_equation} we have determined the unknown distribution functions ($f_1,f_2$) at the wall while satisfying the free-slip condition. Similar strategy is followed to determine the unknown distributions corresponding to the momentum and energy equations as shown below.

For the $x$-momentum equation,
$\rho uv= P\mathbf{\Lambda_2} \mathbf{f^b}=\lambda(-f^b_1-f^b_2+f^b_3+f^b_4)=0$

This condition is satisfied when
\begin{equation}
f^b_1=f^b_4
\end{equation}
\begin{equation}
f^b_2=f^b_3
\end{equation}

For the $y$-momentum equation,
$\rho uv= P\mathbf{\Lambda_1} \mathbf{f^c}=\lambda(-f^c_1+f^c_2+f^c_3-f^c_4)=0$

This condition is satisfied when
\begin{equation}
f^c_1=-f^c_4
\end{equation}
\begin{equation}
f^c_2=-f^c_3
\end{equation}

For the energy equation, we have
$ Pv+\rho vE= P\mathbf{\Lambda_2} \mathbf{f^d}=\lambda(-f^d_1-f^d_2+f^d_3+f^d_4)=0$

This condition is satisfied when
\begin{equation}
f^d_1=f^d_4
\end{equation}
\begin{equation}
f^d_2=f^d_3
\end{equation}

\section{Results and Discussion}
In this section, results obtained by LBRS for several benchmark test cases for inviscid compressible flow are presented.  We have included results for both hyperbolic scalar conservation laws (Burgers equation) in 1-D and 2-D and hyperbolic vector systems (Euler equations or equations of inviscid compressible flows) in 1-D and 2-D.  

\subsection{Sod's shock tube}
In this test case, gas at two different pressures is separated by a diaphragm. Upon rupturing of the diaphragm, an unsteady flow consisting of a moving shock, an evolving simple centered expansion fan and a moving contact discontinuity is established. The initial conditions for this test case are\\

 $ for \> x<0,\quad \left[{\begin{array}{c}
                   \rho_L\\
                    u_L\\
                    p_L\\
                  \end{array}}\right]= \left[{\begin{array}{c}
                                        1.0 \> kg/m^3\\
                                        0.0 \> m/s\\
                                        100000 \> N/m^2\\
                                        \end{array}}\right] \>; \qquad for\> x>0, \quad  \left[{\begin{array}{c}
                   \rho_R\\
                    u_R\\
                    p_R\\
                  \end{array}}\right]= \left[{\begin{array}{c}
                                        0.125 \> kg/m^3\\
                                        0.0 \> m/s\\
                                        10000 \> N/m^2\\
                                        \end{array}}\right] \>$\\

\begin{figure}[H]
\centering
\includegraphics[trim = 4mm 4mm 10mm 10mm,clip,width=8cm]{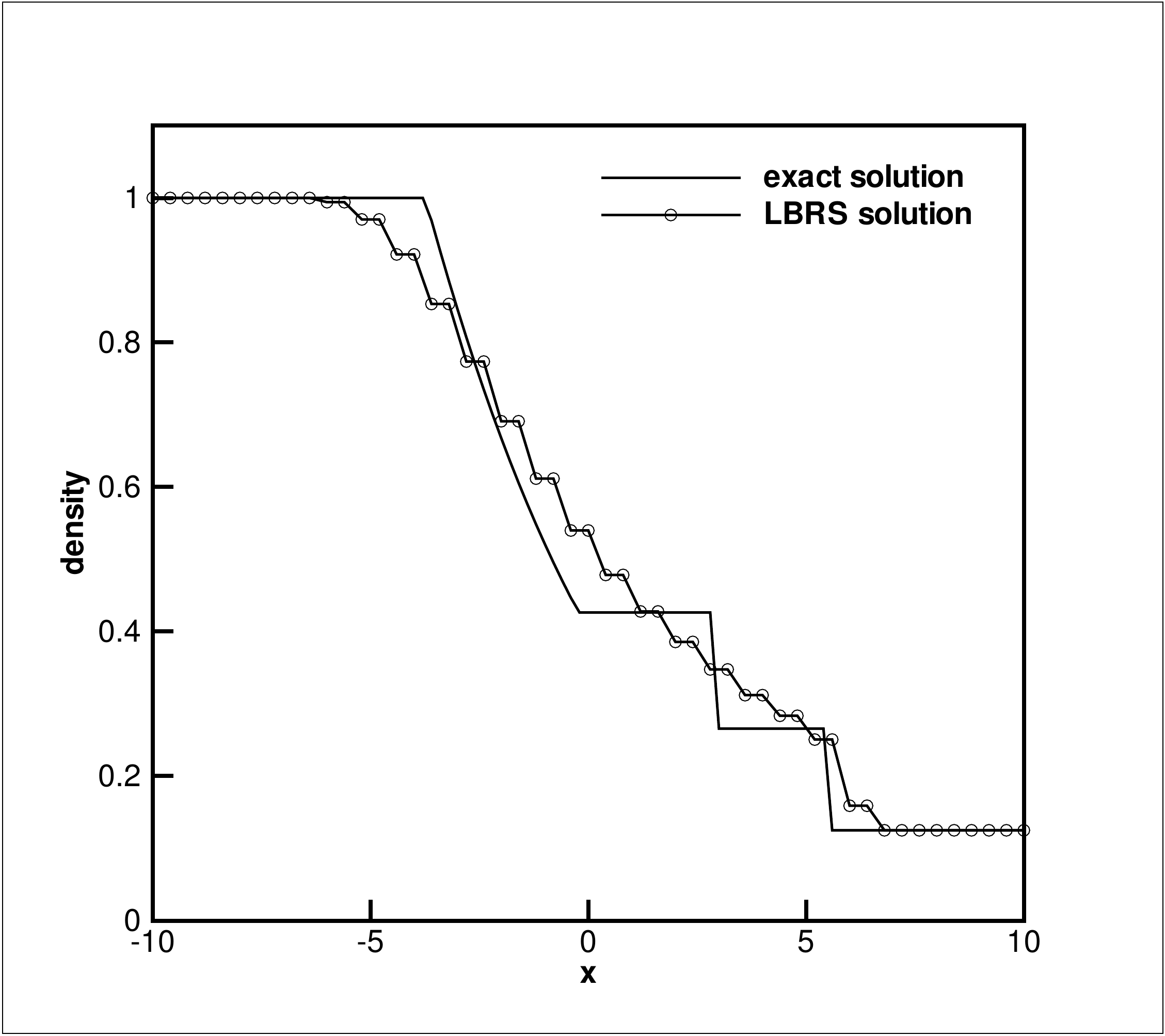}
\caption{Density plot for shock tube test case for the D1Q2 model}
\label{fig:ShocktubeD1Q2}
\end{figure}

\begin{figure}[H]
\centering
\includegraphics[trim = 10mm 5mm 10mm 10mm,clip,width=8cm]{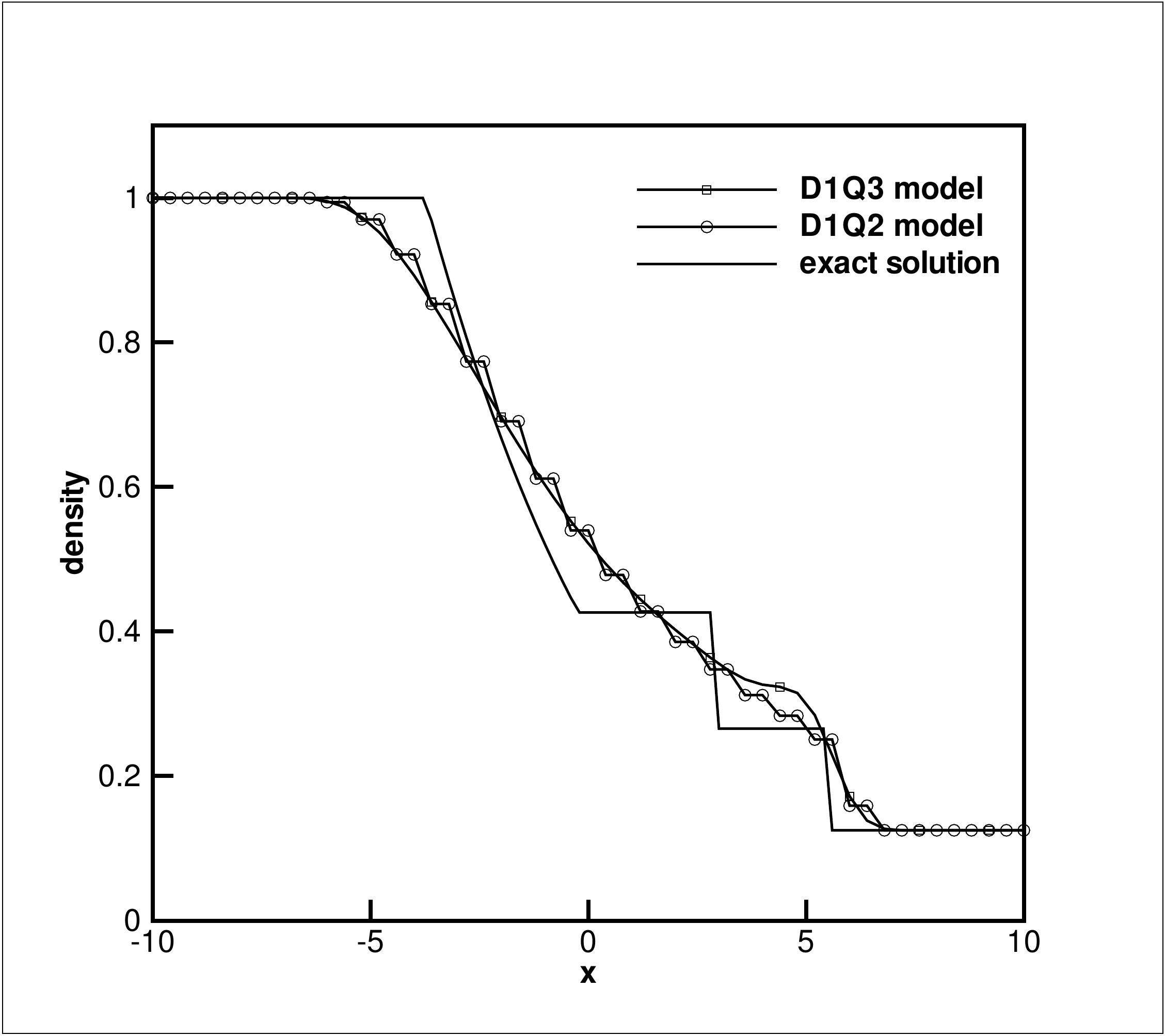}
\caption{Comparison of shock tube test case results for the D1Q3 and D1Q2 models}
\label{fig:shocktubeD1Q3}
\end{figure}

We have divided the domain [-10, 10] into 50 cells. The results are presented at time $t=0.01$ seconds. In Figure (\ref{fig:ShocktubeD1Q2}), the density plot obtained by the D1Q2 model is compared with the analytical solution.  The result for the D1Q2 model exhibits a staircase-like structure which is similar to the result produced by Lax-Friedrichs method in CFD.  With the addition of a rest particle, the D1Q3 model however gives smooth results (see figure (\ref{fig:shocktubeD1Q3})).   Thus the three-velocity model is an improvement over the two-velocity model, though both simulate the compressible flow and resolve all the flow features involved.  However, we can conclude from the Sod\textquoteright s Shock tube test case results that both the models involve significant numerical diffusion.  This is expected because the number of equations is typically more than the number of equations in the original conservation laws and numerical diffusion gets augmented for each equation.  It may be possible to control this numerical diffusion in the LBRS framework but this is beyond the scope of the present work and is a possibility for future research.    

\subsection{Steady Shock}
\begin{figure}[H]
\centering
\includegraphics[trim = 10mm 5mm 10mm 10mm,clip,width=8cm]{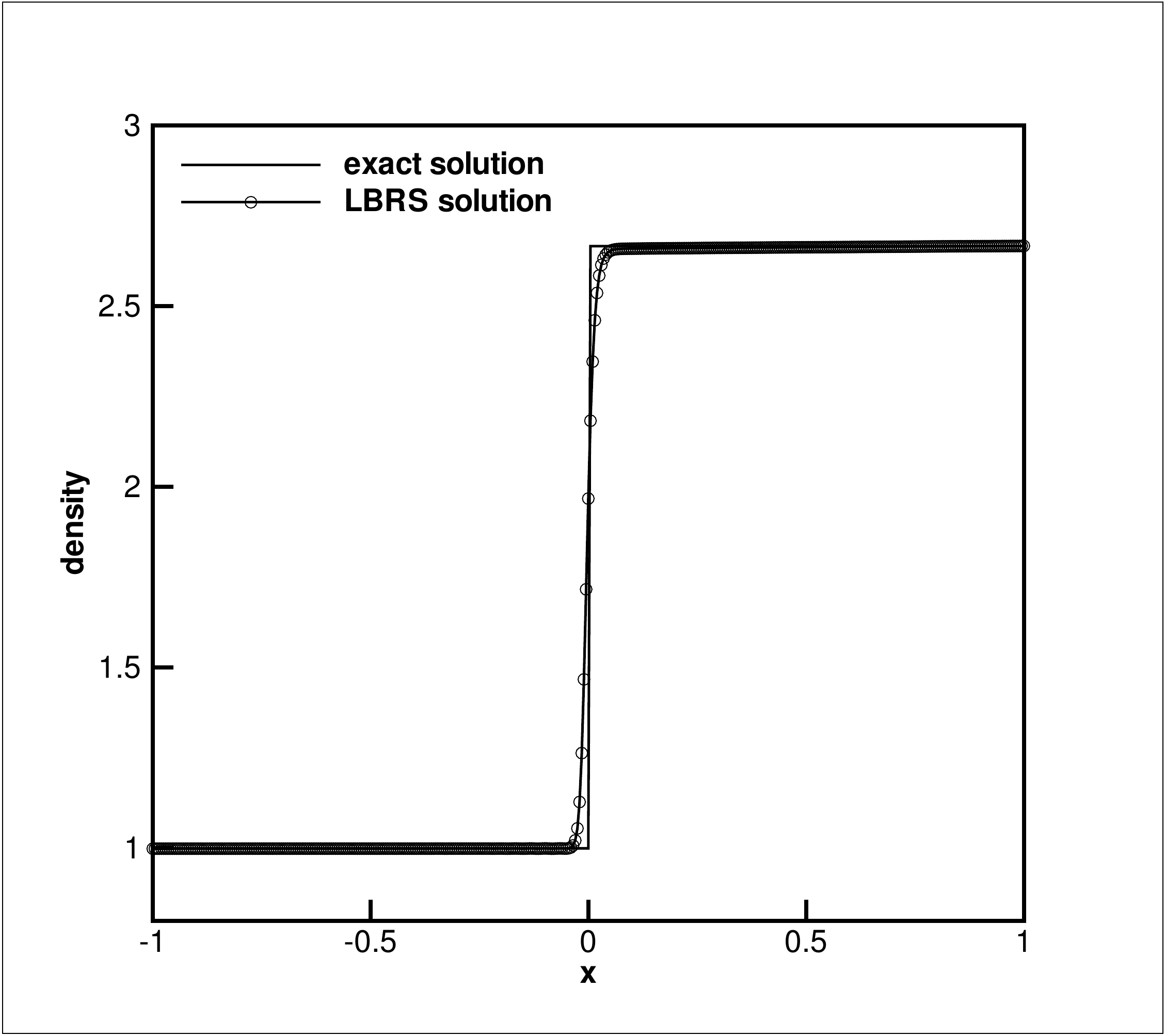}
\caption{LBRS result for the steady shock test case}
\label{fig:steadyshock}
\end{figure}

The next test case presents the LBRS solution for a steady shock in a shock tube for the 1-D Euler equations.  This test case is taken from Zhang and Shu~\cite{Zhang_Shu}.  The computational domain [-1, 1] is divided into 400 grid points and the shock is located at $x = 0$. The LBRS result is shown in figure (\ref{fig:steadyshock}). The initial Mach number to the left of the shock is equal to 2 and the post-shock conditions are determined from the Rankine-Hugoniot relations. LBRS (with D2Q3 model) captures the steady shock quite well. 

\subsection{Steady contact}
\begin{figure}[H]
\centering
\includegraphics[trim = 10mm 5mm 10mm 10mm,clip,width=8cm]{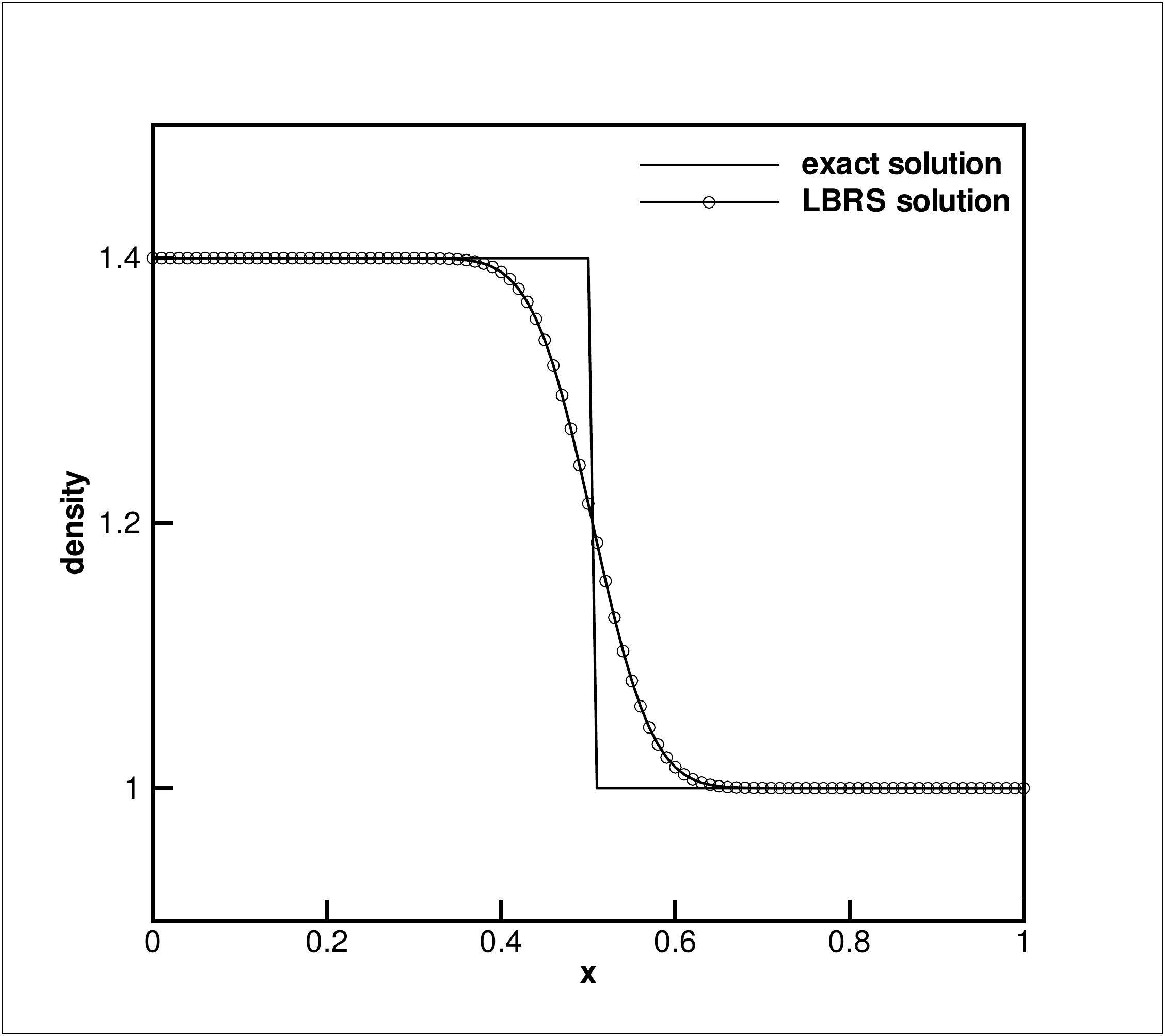}
\caption{LBRS result for the steady contact discontinuity test case}
\label{fig:steadycontact}
\end{figure}

We tested our scheme on the steady contact test case~\cite{Toro} for the 1-D Euler equations. The initial conditions for this test case are $\> \rho_L = 1.4, \> u_L = 0.0, \> p_L = 1.0$ for $x < 0$ and  $\> \rho_R = 1.0, \> u_R = 0.0, \> p_R = 1.0$ for $x > 0$. The domain [0, 1] is divided into 100 grid points and the solution is computed at time $t = 2$. Figure (\ref{fig:steadycontact}) presents the result with LBRS for this test case.   The contact discontinuity is captured reasonably well, though there is significant numerical diffusion present.      

\subsection{Two-dimensional scalar conservation law}
We now present results for a two-dimensional scalar conservation law using the D2Q4 model. We discuss two test cases from Spekreijse~\cite{Spekreijse}. The governing equation for these two test cases is the 2-D inviscid Burgers equation:
\begin{equation}
\frac{\partial{u}}{\partial{t}}+\frac{\partial{(u^2/2)}}{\partial{x}}+\frac{\partial{u}}{\partial{y}}=0
\end{equation}

\begin{figure}[H]
\centering
\includegraphics[trim = 10mm 5mm 10mm 10mm,clip,width=6cm]{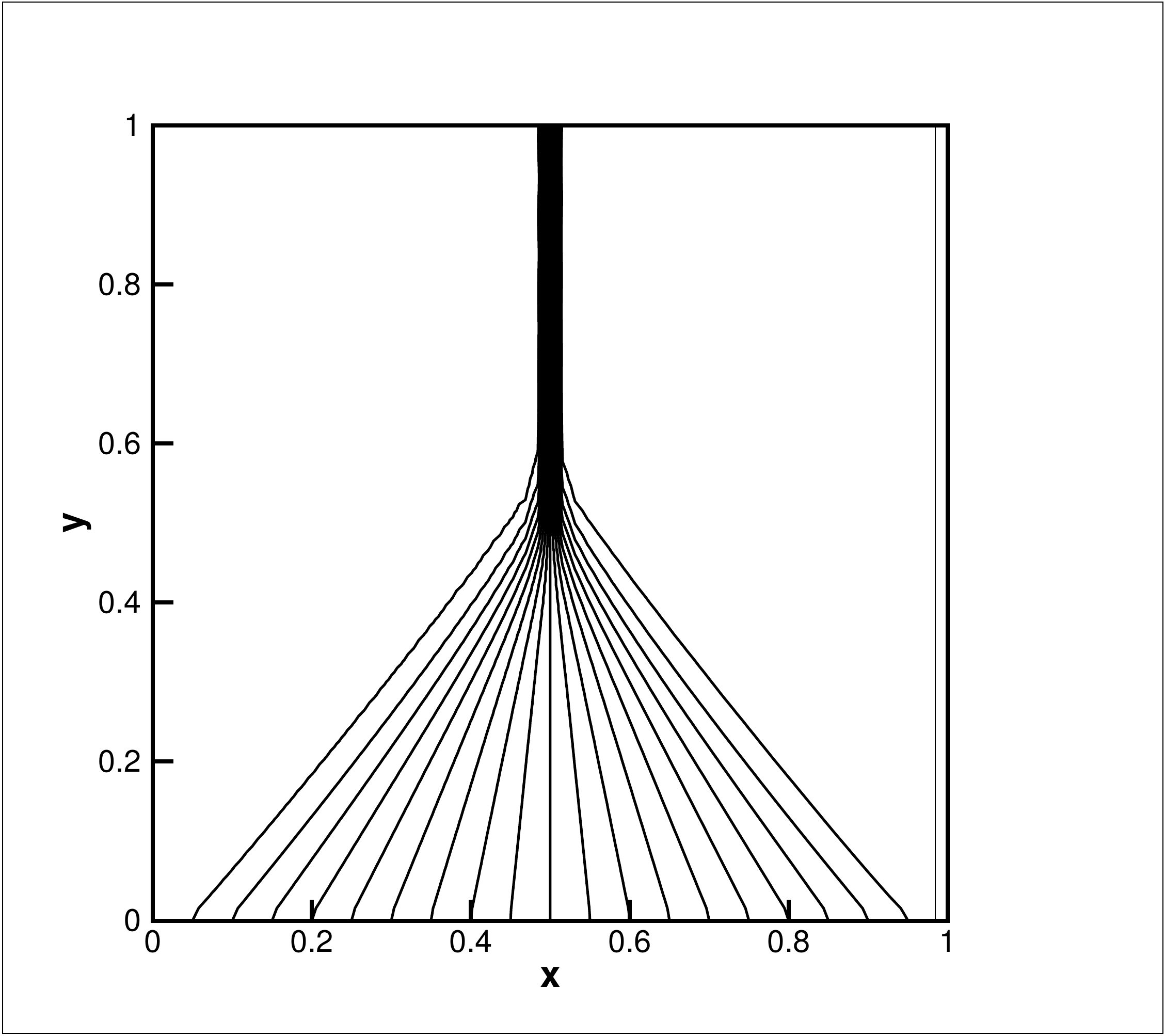}
\caption{density plot for case 1 of the two-dimensional Burgers equation}
\label{fig:normal shock case}
\end{figure}

The problem is solved in the domain [0, 1] x [0, 1] on a 64 x 64 grid. The boundary conditions for the first case are:
$ \quad u(0, y)=1 \>;\quad u(1, y)=-1 \>; \quad u(x, 0)=1-2x$\\
These boundary conditions result in a normal shock originating at $x=0.5, \>y=0.5$ and a smooth variation resembling an expansion fan below the shock.  Figure (\ref{fig:normal shock case}) shows that both these features are captured well by LBRS.  

\begin{figure}[H]
\centering
\includegraphics[trim = 10mm 5mm 10mm 10mm,clip,width=6cm]{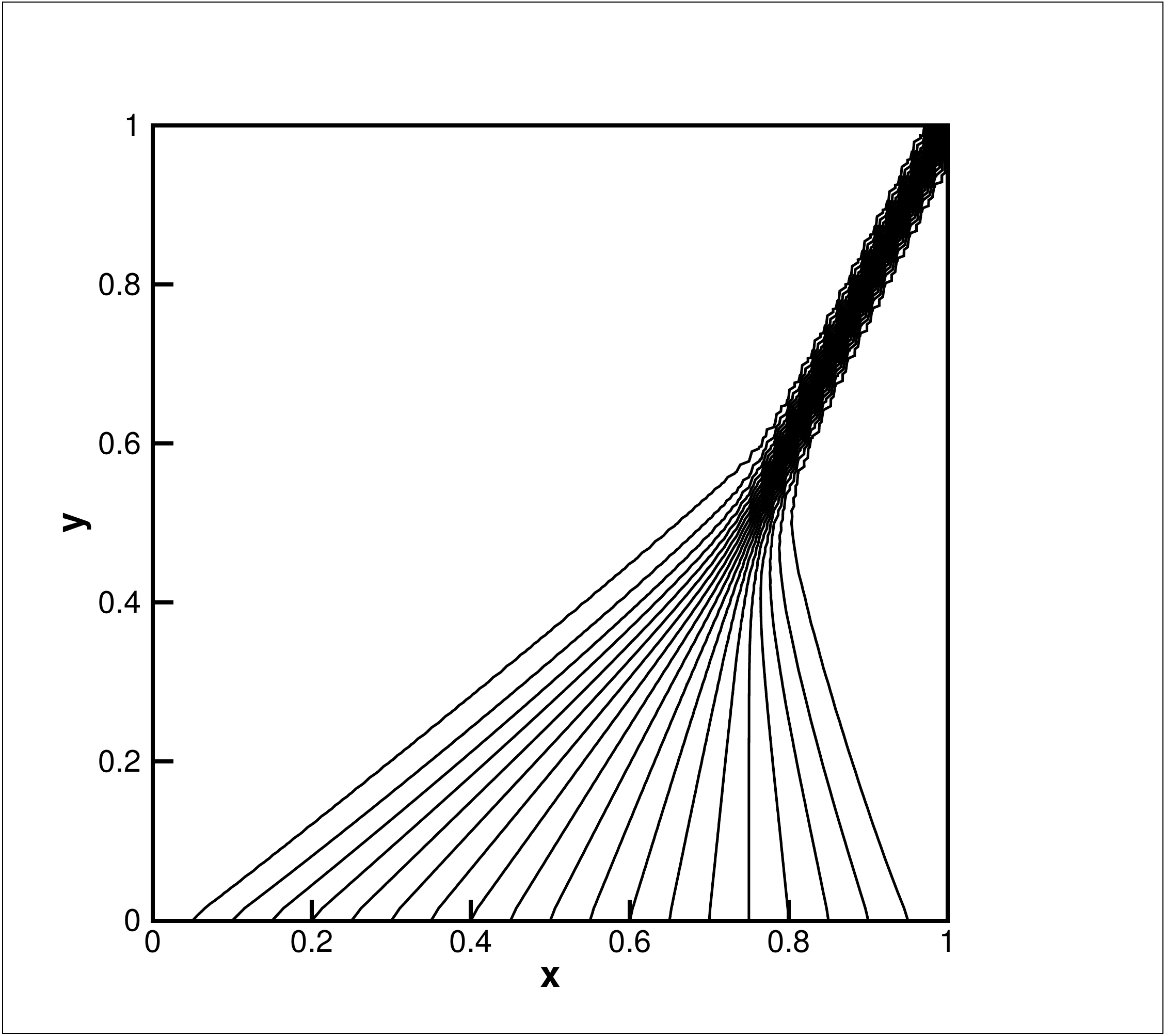}
\caption{density plot for case 2 of the two-dimensional Burgers equatuion}
\label{fig:oblique shock case}
\end{figure}

The boundary conditions are altered for the second test case such that an oblique shock occurs in the top-right corner a shown in figure (\ref{fig:oblique shock case}) instead of a normal shock. The conditions at the boundary for this case are:\\
$\quad u(0, y)=1.5 \>; \quad u(1, y)=-0.5 \>; \quad u(x, 0)=1.5-2x$\\
Again, both the shock and the smooth variation are captured well by LBRS.  

After demonstrating that our method works satisfactorily for a scalar two-dimensional conservation law, we now present results for the two-dimensional Euler equations.
\subsection{Oblique shock reflection test case}
\begin{figure}[H]
\centering
\includegraphics[trim = 10mm 1mm 10mm 2mm,clip,width=14cm]{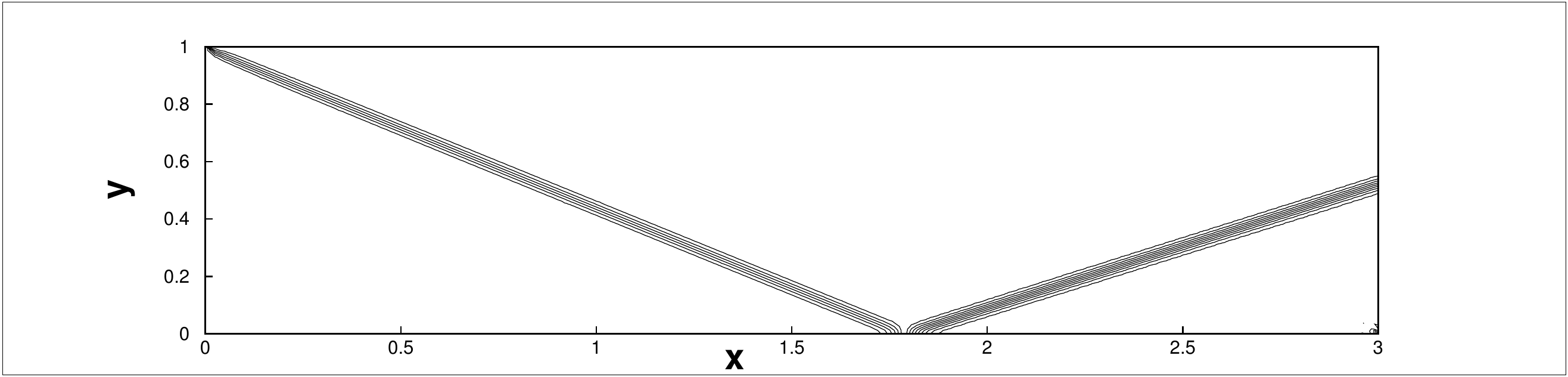}
\caption{Density contours for the oblique shock reflection test case on a 240x80 grid}
\label{fig:oblique shock reflection}
\end{figure}

This test case deals with the reflection of an oblique shock wave from a solid wall. The boundary conditions at the inflow and the top boundary (post-shock boundary conditions) are set such that an oblique shock enters the domain from the top-left boundary. This shock is then reflected from the bottom wall. The values of the primitive variables at the inflow boundary are $\> \rho=1.0, \> u=2.9, \> v=0.0, \> P=1.0/1.4$ and the corresponding post-shock values at the top boundary are given by $\> \rho=1.69997, \> u=2.61934, \> v=-0.50633, \> P=1.52819$, calculated from the oblique shock relations from gas dynamics.  The free-slip boundary condition discussed in the previous section is applied at the solid wall. The density contours obtained using LBRS are plotted in figure (\ref{fig:oblique shock reflection}).  It can be seen from the figure that LBRS captures both the incident and the reflected shocks satisfactorily.  

\subsection{Explosion problem}
We now present results for the explosion problem which can be considered as a two-dimensional extension of the Sod\textquoteright s shock tube problem\cite{Toro}. The computational domain is a square of dimensions [-1, 1] x [-1, 1] and is divided into 400 x 400 cells. The initial conditions are\\
$$\quad \left[{\begin{array}{c}
             \rho\\
             u\\
             v\\
             p\\
             \end{array}}\right]=\left[{\begin{array}{c}
                                        1.0\\
                                        0.0\\
                                        0.0\\
                                        1.0\\
                                        \end{array}}\right] \> for \> x^2+y^2<0.4^2 \>; \qquad \qquad \left[{\begin{array}{c}
                 \rho\\
                    u\\
                    v\\
                    p\\
                 \end{array}}\right]=\left[{\begin{array}{c}
                                      0.125\\
                                      0.0\\
                                      0.0\\
                                      0.1\\
                                      \end{array}}\right] \>for \> x^2+y^2>0.4^2 $$ 

\begin{figure}[H]
\centering
\includegraphics[trim = 10mm 10mm 10mm 10mm,clip,width=12cm]{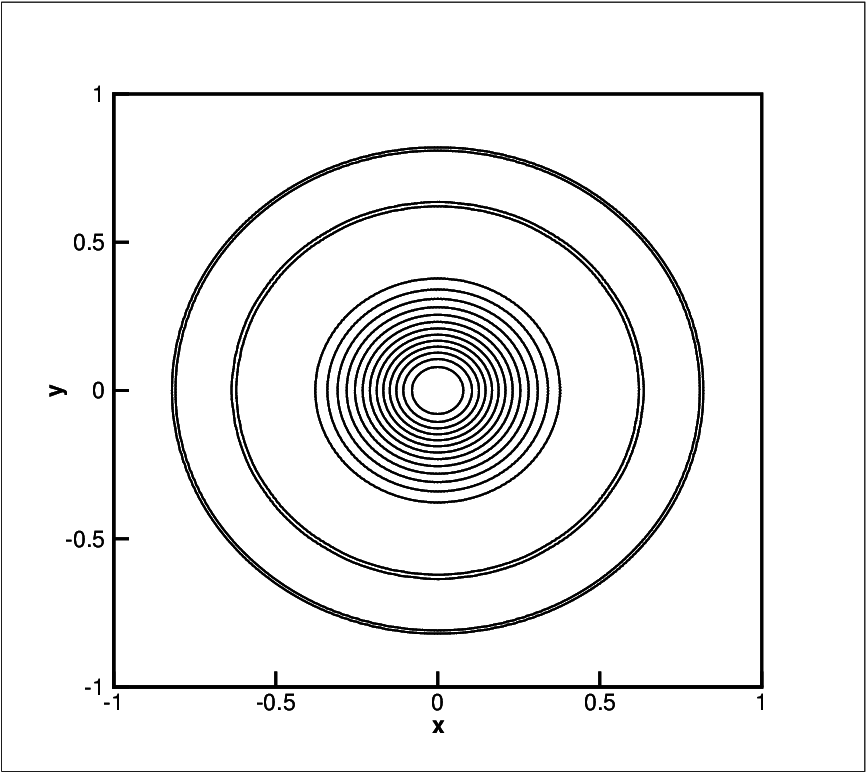}
\caption{Density contours for the explosion case}
\label{fig:cylinder explosion}
\end{figure}

\begin{figure}[H]
\centering
\includegraphics[trim = 10mm 5mm 10mm 10mm,clip,width=8cm]{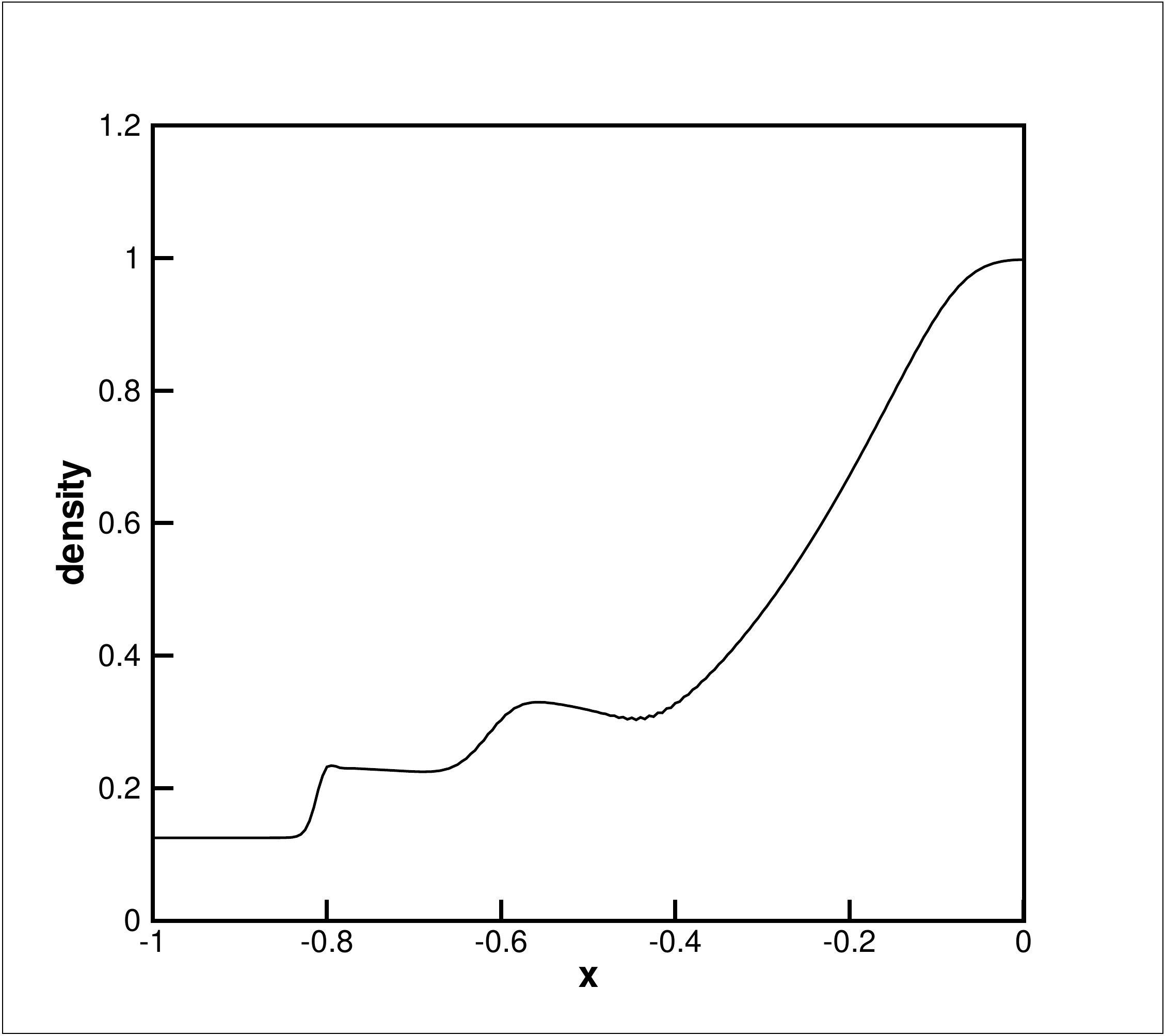}
\caption{Density along the centreline $y=0$ for the explosion case}
\label{fig:density explosion}
\end{figure}

\begin{figure}[H]
\centering
\includegraphics[trim = 10mm 5mm 10mm 10mm,clip,width=8cm]{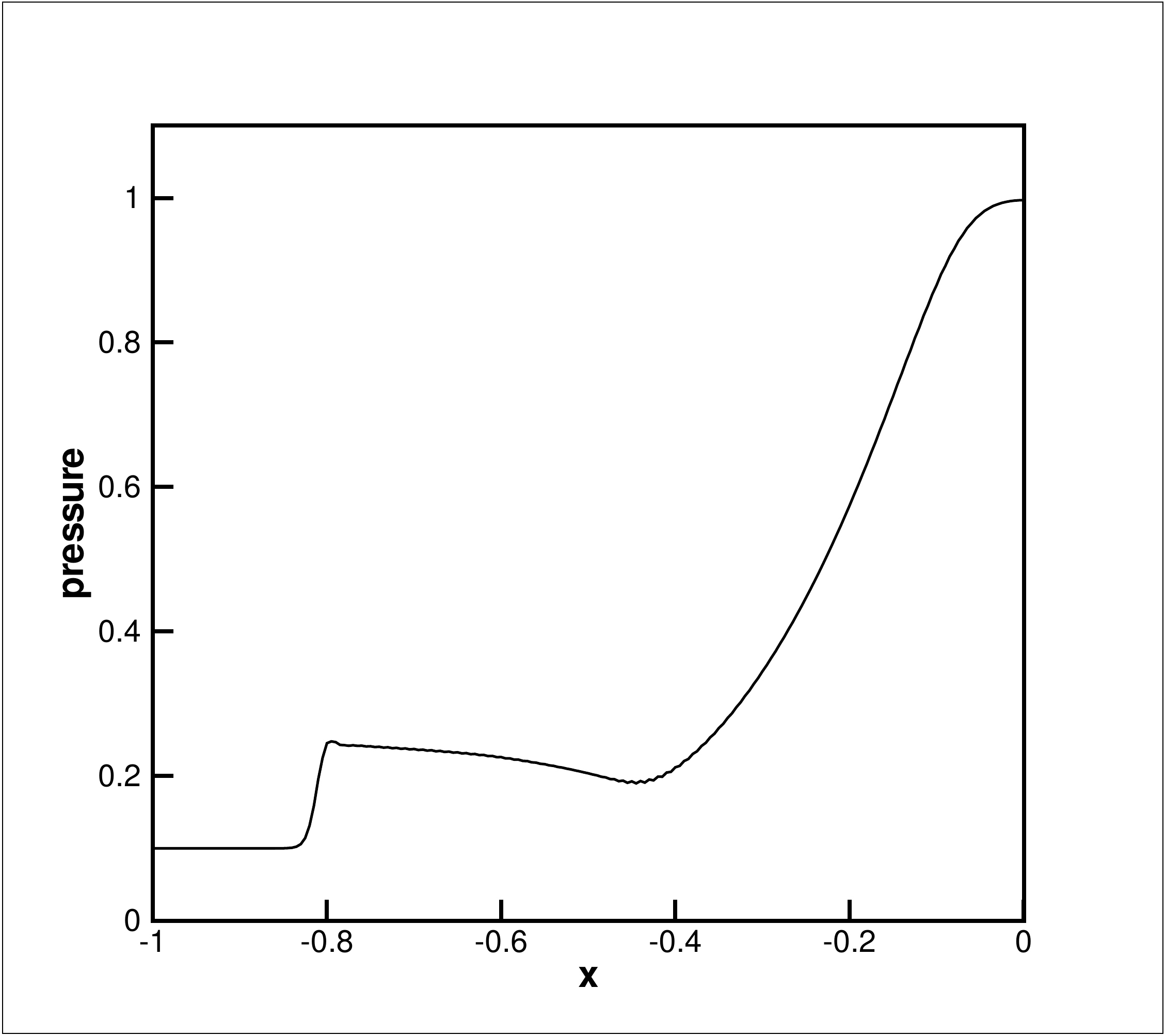}
\caption{Pressure along the centreline $y=0$ for the explosion case}
\label{fig:pressure explosion}
\end{figure}

We plot the results for time $t=0.25$.  At this time, the solution consists of a circular shock wave traveling away from the origin, a circular contact surface traveling in the same direction as the shock and a circular expansion wave traveling inwards towards the origin.   LBRS captures all the three waves accurately (with respect to position) as shown in figures (\ref{fig:cylinder explosion}), (\ref{fig:density explosion}) and (\ref{fig:pressure explosion}).  

\subsection{2-D Riemann problem}
The two-dimensional Riemann problem is solved on the square [0, 1] x [0, 1]. The domain is divided into four quadrants, each of them having different initial conditions. The initial conditions are:\\
$\> \left[{\begin{array}{c}
             \rho\\
             u\\
             v\\
             p\\
             \end{array}}\right]=\left[{\begin{array}{c}
                                        1.1\\
                                        0.0\\
                                        0.0\\
                                        1.1\\
                                        \end{array}}\right] \> for \> x\geq 0,\> y\geq 0 \>; \quad \left[{\begin{array}{c}
                 \rho\\
                    u\\
                    v\\
                    p\\
                 \end{array}}\right]=\left[{\begin{array}{c}
                                      0.5065\\
                                      0.8939\\
                                      0.0\\
                                      0.35\\
                                      \end{array}}\right] \>for \> x<0,\> y>0.0 \>; \\ \left[{\begin{array}{c}
             \rho\\
             u\\
             v\\
             p\\
             \end{array}}\right]=\left[{\begin{array}{c}
                                        1.1\\
                                        0.8939\\
                                        0.8939\\
                                        1.1\\
                                        \end{array}}\right] \> for \> x\leq 0,\> y\leq 0 \>; \quad \left[{\begin{array}{c}
                 \rho\\
                    u\\
                    v\\
                    p\\
                 \end{array}}\right]=\left[{\begin{array}{c}
                                      0.5065\\
                                      0.0\\
                                      0.8939\\
                                      0.35\\
                                      \end{array}}\right] \>for \> x>0,\> y<0.0 \>$\\

\begin{figure}[H]
\centering
\includegraphics[trim = 10mm 10mm 10mm 5mm,clip,width=8cm]{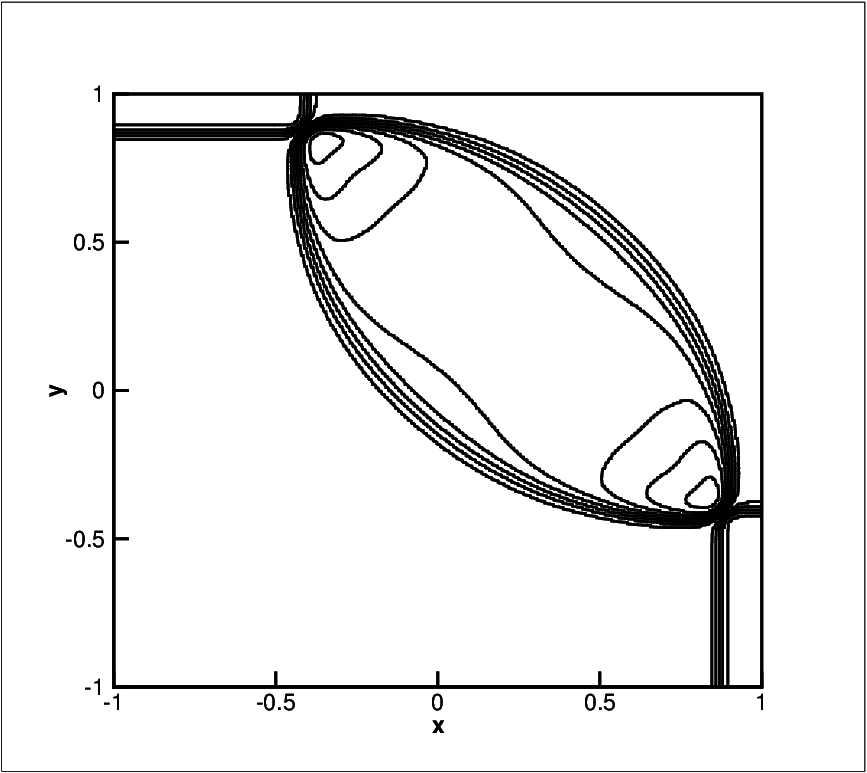}
\caption{Density contours for the 2-D Riemann problem on a 400 x 400 grid}
\label{fig:Riemann_problem}
\end{figure}

The test case presented here corresponds to case 4 in Liska and Wendroff~\cite{Liska_Wendroff}.  At each interface, we have a one-dimensional Riemann problem.  The initial conditions are selected in such a way that a single wave (either a shock, contact-slip or a rarefaction wave) is produced  for each 1-D Riemann problem.  For the given initial conditions, the solution consists of four shock waves evolving in the domain, two of which are straight shocks while the other two are curved shocks which border the lens-shaped region (see figure (\ref{fig:Riemann_problem})).  The LBRS result captures these flow features well as can be seen from the figure (\ref{fig:Riemann_problem}).   Liska and Wendroff report that the solution must be symmetric about the axis of the lens and our scheme maintains this symmetry.

\subsection{Supersonic flow over a forward-facing step}
\begin{figure}[H]
\centering
\includegraphics[trim = 10mm 0.8mm 1mm 1mm,clip,width=12cm]{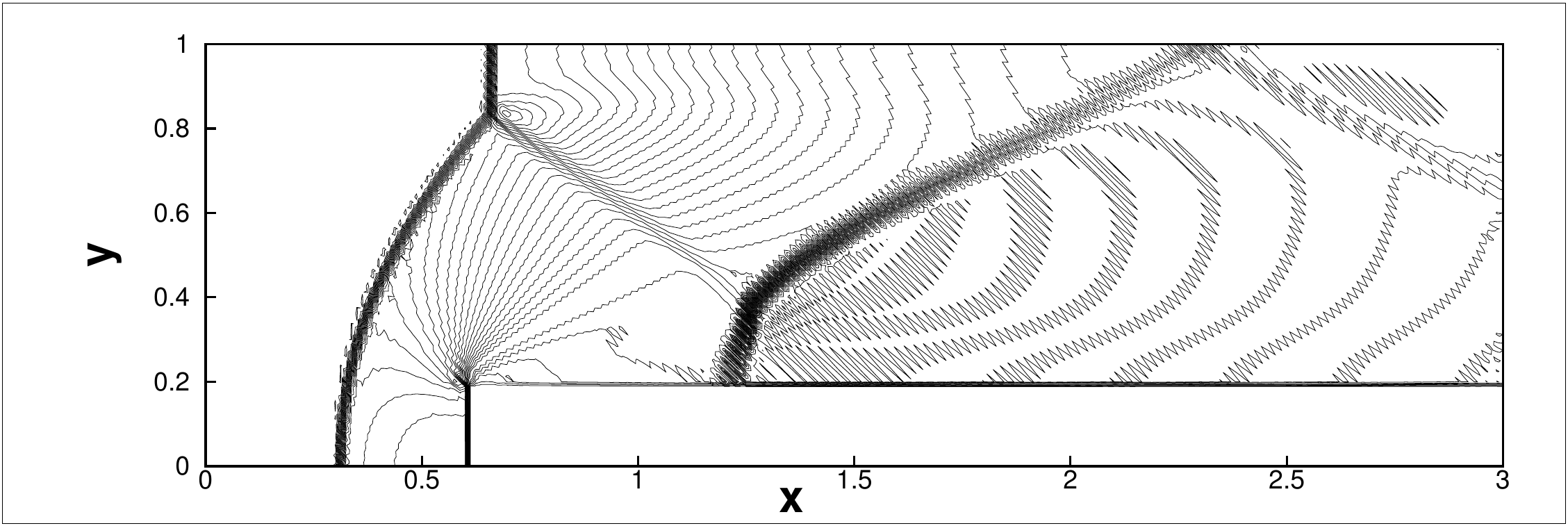}
\caption{Density contours for Mach 3 flow over a forward-facing step on a 240x80 grid}
\label{fig:ffs}
\end{figure}

This test case simulates supersonic flow over a forward facing step in a channel.  The channel dimensions are [0, 3] x [0, 1] and the inflow Mach number is 3. The initial conditions are set to the inflow boundary values and the free-slip boundary condition is enforced at the walls. The density contours (see figure (\ref{fig:ffs})) are plotted at time $t=4$.  Though the incident and reflected shocks, the expansion fan originating the tip of the step and the shock-expansion interaction are captured well, the contours for this test case show a lot of non-smoothness.  We noticed this non-smoothness in the case of vector conservation laws and only for some test cases.  An improvement to overcome this non-smoothness is suggested in the conclusion section.  

\subsection{Shock diffraction}
\begin{figure}[H]
\centering
\includegraphics[trim = 10mm 2mm 10mm 5mm,clip,width=8cm]{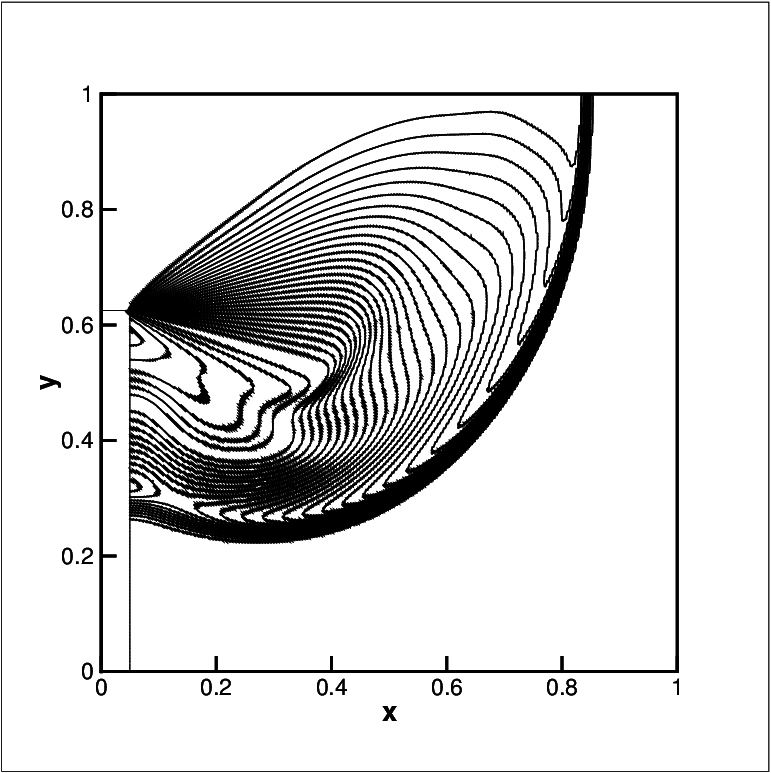}
\caption{Density contours for the shock diffraction test case on a 400x400 grid}
\label{fig:shockdiffraction}
\end{figure}

In this test case, a moving normal shock (Mach number 5.09) encounters a backward-facing step.  Initially, the normal shock is placed along the tip of the step.  The shock wave diffracts as it passes over the step. The computational domain for this case is a unit square [0, 1] x [0, 1]. The LBRS simulation is shown in figure (\ref{fig:shockdiffraction}).   The density contours are plotted at time $t=0.1561$.  Here too, the contours are non-smooth but the flow features are all resolved well.  

\subsection{Supersonic flow over a compression ramp}
\begin{figure}[H]
\centering
\includegraphics[trim = 10mm 0.8mm 10mm 5mm,clip,width=12cm]{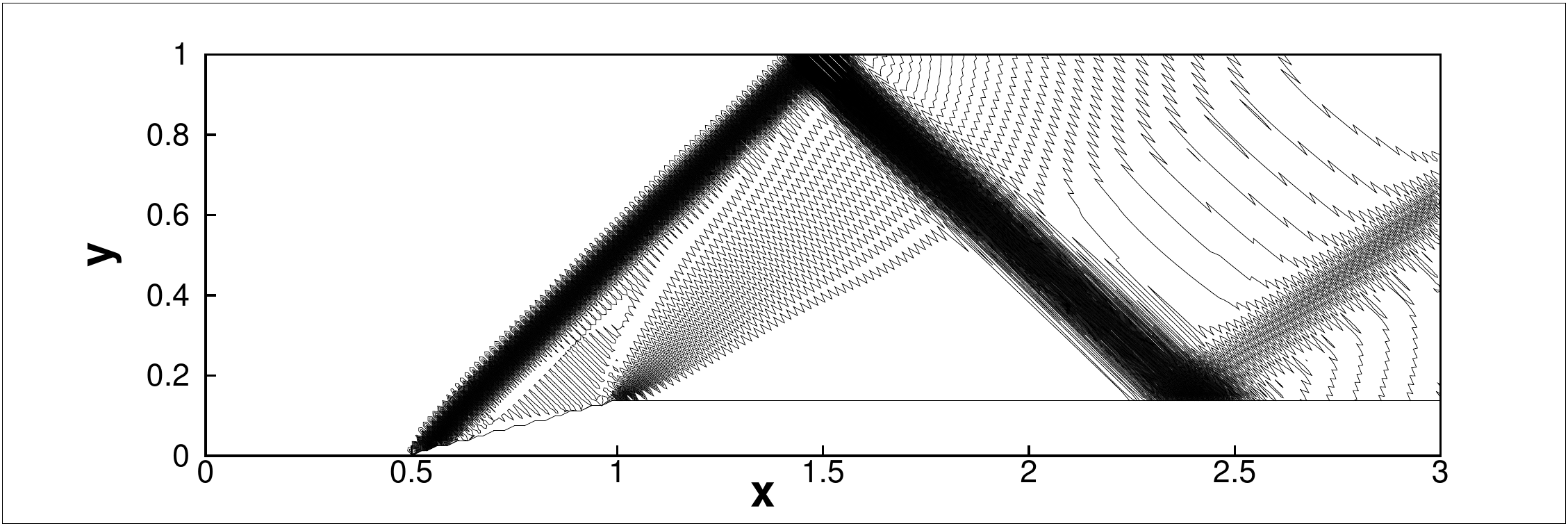}
\caption{Density contours for supersonic flow over a ramp in a channel on a 240x80 grid}
\label{fig:ramp}
\end{figure}

Result for this test case involving supersonic flow (M=2) over a compression ramp is shown in the (figure \ref{fig:ramp}).  Here too, the flow features involving initial and reflected oblique shocks, expansion wave and shock-expansion interaction are captured well but the non-smoothness in the contours appear again.  

We have presented results for a variety of test cases simulating inviscid compressible flows to showcase the capability of LBRS in resolving flow features.   Our scheme successfully captures the flow features involved in all the test cases.  However, the results for some of the cases for two-dimensional flows, for vector conservation laws, show non-smoothness in the contours.  We guess that as in 1-D where addition of more discrete velocities improved the results, increasing discrete velocities may solve the problem of non-smoothness.  Numerical experiments with a D2Q9 model will be presented in a forthcoming paper \cite{Rohan_SVRRao_LBRS}, together with several additional simulations for compressible flows with curved bodies.   

\section{Conclusion}
A simple method based on the interpretation of the diagonal form of the relaxation system as a discrete velocity Boltzmann equation and further utilization of this relaxation system as a lattice Boltzmann equation has been presented in this work.  This novel lattice Boltzmann relaxation scheme (LBRS)  does not involve any low Mach number expansion of the equilibrium distribution functions as in the traditional LBM.  The equlibrium distribution functions in this method are simple algebraic functions of the conserved variable and the fluxes.  Thus, the LBRS can successfully simulate compressible fluid flows.   We have tested our scheme on a variety of standard test cases both for one and two dimensions and for both scalar and vector conservation laws, namely for the inviscid Burgers equation and the Euler equations of compressible fluid flows in both 1-D and 2-D. LBRS successfully captures all the flow features like shock waves, contact discontinuities and expansion waves.  In the 2-D cases for some of the test cases for vector conservation laws, non-smoothness in the contours is observed and an improved model to over come this drawback will be presented in a future work \cite{Rohan_SVRRao_LBRS} together with more simulations for compressible flows with curved boundaries.


\begin{thebibliography}{99} 
\bibitem{Chen_Doolen} S. Chen and G.D. Doolen, Lattice Boltzmann method for fluid flows, Annu. Rev. Fluid Mech., 30, 329-364, 1998.  

\bibitem{Benzi_Succi_Vergasola} R. Benzi, S. Succi and M. Vergassola, The lattice Boltzmann equation: Theory and Applications, Physics Reports, 222, no.3, 145-197, 1992.  

\bibitem{Rothman_Zaleski} D.H. Rothman and S.~Zaleski, Lattice-Gas Cellular Automata: Simple Models of Complex Hydrodynamics, Cambridge University Press, 1997.

\bibitem{Wolf-Gladrow} Dieter~A. Wolf-Gladrow, Lattice-Gas Cellular Automata and Lattice Boltzmann Models: An Introduction, Springer-Verlag, 2000.

\bibitem{Rivet_Boon} J.-P. Rivet and J.P. Boon, Lattice Gas Hydrodynamics,  Cambridge University Press, 2001.

\bibitem{Succi} Sauro Succi, The Lattice Boltzmann Equation for Fluid Dynamics and Beyond, Clearendon Press - Oxford, 2001.  

\bibitem{Sukop_Thorne} Michael~C. Sukop and Jr. Daniel T.~Thorne, The Lattice Boltzmann Modeling: An Introduction for Geoscientists and Engineers, Springer-Verlag, 2006.  

\bibitem{Mohamad} A.A. Mohamad, Lattice Boltzmann Method: Fundamentals and Engineering Applications with Computer Codes, Springer-Verlag, 2011.

\bibitem{McNamara} G. R. McNamara and G. Zanetti, Use of the Boltzmann equation to simulate lattice-gas automata, Phys. Rev. Lett., 61(20), 2332-2335, 1988.  

\bibitem{Higuera} F. J. Higuera and J. Jimenez, Boltzmann approach to latttice gas simulations, Europhys. Lett, 9(7), 663-668, 1989.  

\bibitem{Chen_Chen} H. Chen, S. Chen and W.H. Matthaeus, Recovery of the Navier-Stokes equations using a lattice-gas Boltzmann method, Phys. Rev. A, 45(8), R 5339-R5342, 1992.  

\bibitem{He_Luo} X. He and Li-Shi Luo, A {\em priori} derivation of the lattice Boltzmann equation, Phys. Rev. E, 55(6), R 6333-R6336, 1997.  

\bibitem{Alexander}  F.J. Alexander, H. Chen, S, Chen and G.D. Doolen, Lattice Boltzmann model for compressible fluids,  Phys. Rev. A, 46(4), 1967-1970, 1992.  

\bibitem{Shouxin} H. Shouxin, Y. Guangwu and S. Weiping, A lattice Boltzmann model for compressible perfect gas, Acta Mechanica Sinica, 13(3), 218-226, 1997.  

\bibitem{Guangwu} Y. Guangwu, C. Yaosong and H. Shouxin, Simple lattice Boltzmann model for simulating flows with shock wave, Phys. Rev. E, 59(1), 454-459, 1999.  

\bibitem{Sun} C. Sun, Lattice Boltzmann models for high speed flows, Phys. Rev. E, 58(6), 7283-7287, 1998.  

\bibitem{Kataoka} T. Kataoka and M. Tsutahara, Lattice Boltzmann method for compressible Euler equations, Phys. Rev. E, 69:056702, 1-14, 2004.  

\bibitem{Tsutahara} T. Kataoka and M. Tsutahara, Lattice Boltzmann model for the compressible Navier-Stokes equations with flexible specific heat ratio, Phys. Rev. E, 69:035701(R), 1-4, 2004.  

\bibitem{Tolke} J. Tolke, A thermal model based on the lattice Boltzmann method for low Mach number compressible flows, Journal of Computational and Theoretical Nanoscience, 3, 1-9, 2006.  

\bibitem{Yan} G. Yan, Y. Dong and Y. Liu, An implicit Lagrangian lattice Boltzmann method for the compressible flows, International J. Numer. Meth. Fluids, 51, 1407-1418, 2006.  

\bibitem{Zhang} J. Zhang, G. Yan, X. Shi and Y. Dong, A Lattice Boltzmann Model for the Compressible Euler Equations with Second Order Accuracy, International J. Numer. Meth. Fluids, 60, 95-117, 2009.  

\bibitem{Liu} G. Yan, J. Zhang, Y. Liu, and Y. Dong, A Multi-Energy Level Lattice Boltzmann Model for the Compressible Navier-Stokes Equations, International J. Numer. Meth. Fluids, 55, 41-56, 2007.  

\bibitem{So} R.M.C. So, R.C.K. Leung and S.C. Fu, Modeled Boltzmann Equation and Its Application to Shock-Capturing Simulation, AIAA Journal, 46(12), 3038-3048, 2008.  

\bibitem{Erdembilegt} J.H.B. Erdembilegt, W.-B. Feng and W. Zhang, High Velocity Flow Simulation using Lattice Boltzmann Method with No-Free-Parameter Dissipation Scheme, Journal of Shanghai University (English Edition), 13(6), 454-461, 2009.  

\bibitem{Fu} R.M.C. So, S.C. Fu and R.C.K. Leung, Finite Difference Lattice Boltzmann Method for Compressible Thermal Fluids, AIAA Journal, 48(6), 1059-1071, 2010.  

\bibitem{Qu} K. Qu, C. Shu, and Y.T. Chew, Lattice Boltzmann and Finite Volume Simulations of Inviscid Compressible Flows with Curved Boundary, Advances in Applied Mathematics and Mechanics, 2(5), 573-586, 2010.  

\bibitem{Jin_Xin} S. Jin and Z. Xin, The relaxation schemes for systems of conservation laws in arbitrary space dimensions, Comm. Pure Appl. Math., 48, 235-277, 1995.  

\bibitem{Natalini} R. Natalini, A discrete kinetic approximation of entropy solutions to multidimensional scalar conservation laws, J. Differential Equations, 148, 292-317, 1998.  

\bibitem{Driollet_Natalini} D. Aregba-Driollet and R. Natalini, Discrete kinetic schemes for multidimensional systems of conservation laws, SIAM J. Numer. Anal., 37(6), 1973-2004, 2000.  

\bibitem{Jayaraj} Jayaraj, A Novel Multi-dimensional Relaxation Scheme for Hyperbolic Conservation Laws, Department of Mechanical Engineering, University B.D.T. College of Engineering, Davanagere, India, June, 2006.   

\bibitem{Arun_RaghuramaRao_1} K.R. Arun, S.V Raghurama Rao, M. Luk\'a\v{c}ov\'a -Medvid\textquoteright ov\'a and Phoolan Prasad, A Genuinely Multi-dimensional Relaxation Scheme for Hyperbolic Conservation Laws, In Proceedings of the seventh ACFD Conference, Indian Institute of Science, Bangalore, pages 1029-1039, November 26-30, 2007.  

\bibitem{Arun_RaghuramaRao_2} K. R. Arun, M.~Luk\'a\v{c}ov\'a - Medvi\softd ov\'a, Phoolan Prasad and S.V. Raghurama Rao, A Second Order Accurate Kinetic  Relaxation Scheme for Inviscid Compressible Flows, in Recent Developments in Numerics of Nonlinear Hyperbolic Conservation Laws, (editors) R. Ansorge, R. Bijl, H. Meister, T. Sonar, Notes on Numerical Fluid Mechanics and Multidisciplinary Design, vol. 120, pp. 1-24, Springer-Verlag, Berlin, 2012.    

\bibitem{Bouchut} F. Bouchut, Construction of BGK models with a family of kinetic entropies for a given system of conservation laws, J. Statist. Phys., 95, 113-170, 1999.  

\bibitem{Zhang_Shu} Shuhai Zhang and Chi-Wang Shu, A New Smoothness Indicator for the WENO Schemes and Its Effect on the Convergence to Steady State Solutions, Journal of Scientific Computing, 31(1/2), 273-305, 2007.  

\bibitem{Toro} E. F. Toro, Riemann Solvers and Numerical Methods for Fluid Dynamics, Springer, Third edition, 2009.  
 
\bibitem{Spekreijse} S. Spekreijse, Multigrid solution of monotone second-order discretizations of hyperbolic conservation laws, Mathematics of computation, 49(179), 135-155, 1987.  
 
\bibitem{Liska_Wendroff} R. Liska and B. Wendroff, Comparison of several difference schemes on 1D and 2D test problems for the Euler equations, SIAM Journal on Scientific Computing, 25(3), 995-1017, 2003.  

\bibitem{Rohan_SVRRao_LBRS} Rohan Deshmukh and S.V. Raghurama Rao, A Lattice Boltzmann Relaxation Scheme for Hyperbolic Systems, in preparation.  

\end{thebibliography}
\end{document}